\documentclass{aa}
\usepackage{graphicx}
\begin{document}

\title{NBODY meets stellar population --- The HYDE-PARC Project}
\author{Andrea Borch \inst{1}
\and Rainer Spurzem  \inst{1}
\and Jarrod Hurley  \inst{2} }

\institute{Astronomisches Rechen-Institut, Heidelberg, Germany
\and Swinburne University, Australia}

\abstract{}{
$N$-body simulations give us a rough idea of how the shape of 
a simulated object appears in three-dimensional space. 
From an observational point of view this may give us a misleading picture 
as a stellar system consists of both bright and faint stars. 
The faint stars may be the most common stars in the 
system but the morphological information obtained by observations of an
object may be dominated by the color properties of the bright stars.
Due to dynamical effects, such as energy equipartition, different masses
of stars may populate different regions in the object. Since stars are 
evolving in mass the stellar evolution may also influence the dynamics of the 
system. Hence, if one is interested in simulating what the morphology will look 
like through a telescope, one needs to simulate in addition evolving stars 
and weight them by their luminosity.}
{Therefore we need to combine simulations of the dynamical evolution and
a stellar population synthesis at the same time. For the dynamical evolution
part we use a parallel version of a direct $N$-body code, NBODY6++. This code
also includes stellar evolution. 
We link the parameters from this stellar evolution routine to the BaSeL 2.0
stellar library. This allows us to obtain a spectrum  and colors for each star
in the simulated cluster. We call this the HYDE-PARC project, which means 
\underline{Hy}brid Code for \underline{D}ynamical \underline{E}volution 
and \underline{P}opulation \underline{A}nalysis of Stella\underline{R} 
\underline{C}lusters.}
{We tested our method by simulating globular clusters with up to 50000 stars 
and investigating the integrated colors. For isolated clusters we found results 
assimilable to standard stellar population synthesis codes such as the 
PEGASE code. For clusters in a tidal field we found that the integrated 
colors become relatively bluer due to energy equipartition effects. In the time
shortly before dissolution of the cluster the stellar M/L ratio becomes
lower compared to isolated clusters. We compared the results of our simulations
to integrated spectra of galactic globular clusters. For the cluster
NGC 1851 we found good agreement between simulation and observation.
For extragalactic clusters in M81 and M31 we looked at medium band spectral energy 
distributions 
and found for some clusters also a good agreement.}{}


\maketitle

\section{Introduction}
Globular clusters are among the oldest objects in a galaxy and also among
the oldest objects in the universe. Hence, the
investigation of these objects is crucial for the understanding of
the historical evolution of their host galaxies. Furthermore, these clusters
provide an interesting environment for investigating dense stellar systems.
In these star clusters collisions between stars happen more often than in
other places in the Galaxy, where typical collision timecales of stars would
exceed the lifetime of the Galaxy or even the age of the universe 
(Davies 2002, Portegies Zwart 2000).
In these systems the effects of stellar dynamics \em and \em stellar evolution
are important and interact with each other. For example, stellar evolution
in close binaries takes place differently than in single stars due to their
mutual interaction, which may cause mass transfer. On the other hand,
stellar evolution changes the stellar mass, and therefore influences the
dynamical evolution of the system. \\

Local globular clusters in our galaxy can be resolved in single stars
and therefore investigated in a detailed manner, for example in terms of 
color-magnitude diagram (CMD) morphologies. 
For more distant extragalactic 
globular clusters both the resolution and the faintness are an issue, even when
 using contemporary instruments such as the Hubble Space Telescope (HST) 
and 8m class telescopes.
The amount of information available for those objects is comparable to that
obtained for clusters in local group galaxies in the early 1990s
(Forbes 2002). Since one has to deal with unresolved objects, integrated
properties such as spectra and colors of the whole stellar population
are available. Therefore one needs to find a way how to derive as much 
information as possible by just looking at integrated spectra or integrated
colors. Due to the combination of dynamical evolution and stellar evolution
mentioned above this task is more complicated than, for example, investigating
the integrated spectra of galaxies. \\

Examples of more recent observations are the spectroscopy of globular clusters
in M81 by Schroder et al. (2002), and the medium-band 
spectral energy distributions (SEDs) now available for globular clusters
in M81 (Ma et al. 2006a) and in M31 (Ma et al. 2006b). 
Narbutis et al. (2006) presented UBVRI broad band photometry of 51 compact
star clusters in M31.
Also, spectra of the NGC 1399 globular clusters are published by 
Kissler-Patig et al. (1998) 
and those of the NGC 4365 clusters are published by Larsen et al. (2001). 
Kundu et al. (2005) did deep HST observations of the cluster system of 
NGC 4365 and NGC 1399, and found that NGC 4365 has a number of globular 
clusters with  bluer optical colors than expected for their red 
optical-to-near-infrared colors. In terms of color, a bimodality of the 
color distribution of globular cluster systems is well-known. The first 
statistical test for bimodality of elliptical galaxy globular cluster colors 
was already presented by Zepf \& Ashman (1993). \\

In order to deal with such problems, one may consider two different 
approaches: (i) follow the evolution of the stars; or, 
(ii) follow the dynamical evolution of the stellar system. \\

The first approach is that of stellar population synthesis models. These 
models follow the evolution of a population of stars. Stars are assumed 
to be born with
a certain initial mass function and a certain star formation rate, which may 
vary in time. Evolutionary models of stars then follow the evolution tracks of
stars in the CMD according to their mass.
Chemical enrichment is also taken into account, because the yields of
supernova explosions will enrich the interstellar medium with metals.
Therefore, the next generation of stars will have a higher metallicity.
The stellar parameters of these stars are transformed into spectra and
colors by using stellar libraries. The summation of the spectra of all
stars lead to integrated properties of the whole stellar population
such as integrated spectra and integrated colors.
There are various codes which follow (in principle) this approach, among 
them there are those of Fioc \& Rocca-Volmerange (1997), 
Bruzual \& Charlot (2003), Anders \& Fritze-v. Alvensleben (2003) 
and Maraston (2005). Recently, Fagiolini et al. (2006) published a Monte Carlo 
simulation of star clusters and addressed the issue of populating stars 
randomly in stellar clusters with limited star numbers. \\

The second type of models are $N$-body simulations where a sample of $N$ particles
interact with each other by gravity. Direct $N$-body simulations follow the 
dynamical evolution of such a sample of particles with high precision.
For each particle, which may be regarded as a ``star'', the trajectory
in space will be followed by calculating the force due to the mutual 
interaction with other stars. In dense stellar systems such as globular 
clusters the dynamical behavior is therefore strongly influenced by the
gravitational interaction with all other stars. As mentioned before, the 
dynamical evolution may be influenced by the evolution of the stars themselves.
During its evolution a star may change in mass, and therefore its dynamical
behavior depends on this mass change. Stellar evolution may take place
differently in double stars. Direct $N$-body codes can also follow the dynamical
evolution of double stars, which are important for the dynamical behavior
of the whole system. \\  

\begin{figure*}
\centering
\includegraphics[width=8cm]{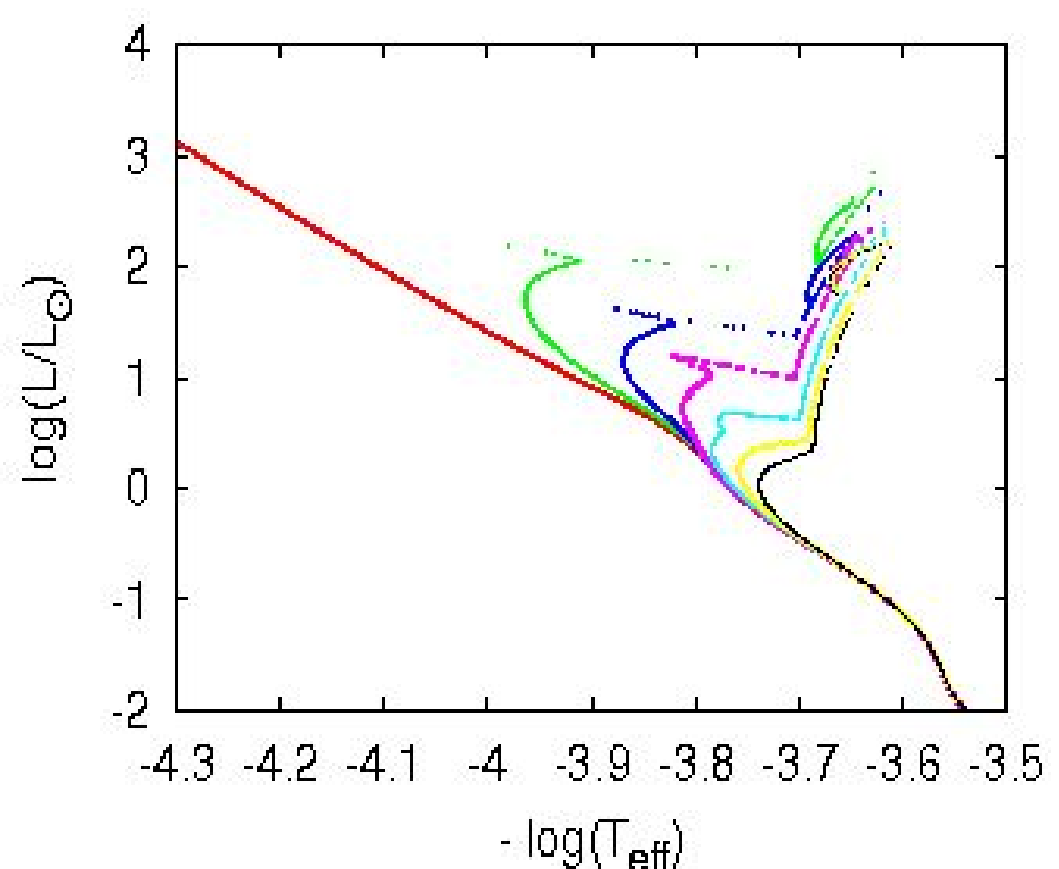}
\includegraphics[width=8cm]{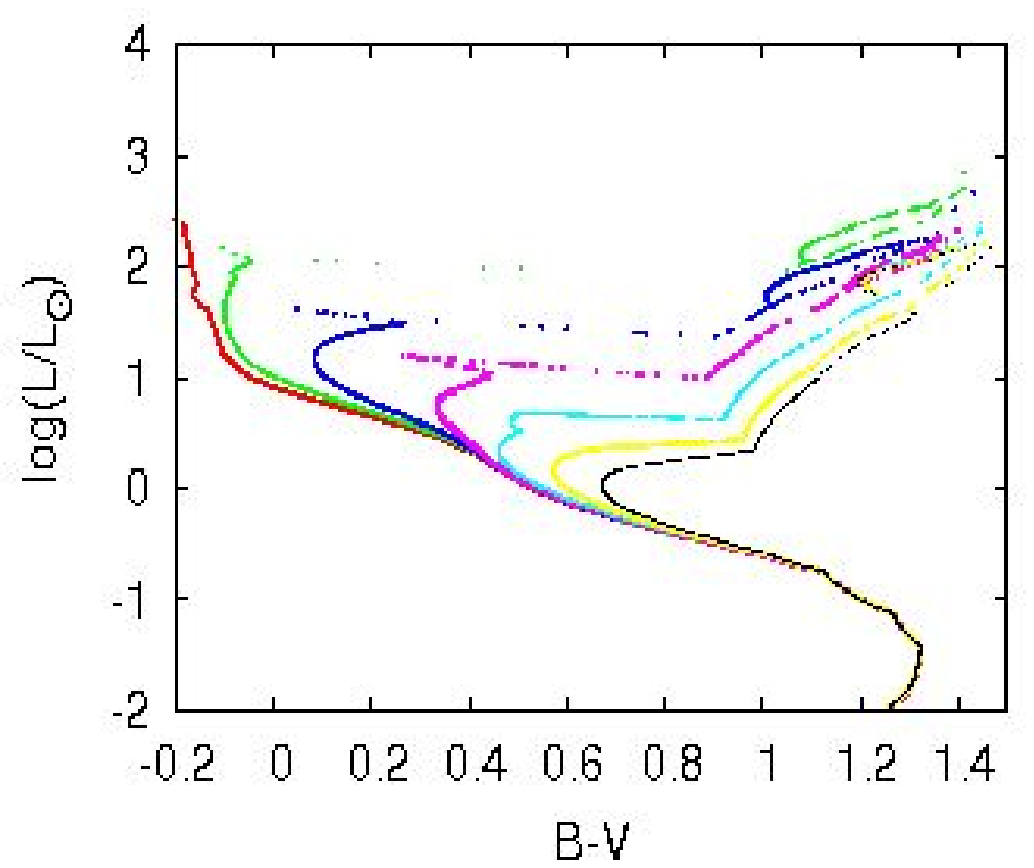}
\caption{Color-magnitude diagram for a simulation with 50000 particles.
The left plot shows the effective temperature versus luminosity. 
The parameters $T_{\rm eff}$ and $\log g$ are drawn from the stellar parameters
 mass, radius and luminosity, delivered by the {\tt hrplot} routine
of the NBODY6++ code. These stellar parameters are transformed into 
spectra and colors by using the stellar library BaSeL.2. The right diagram
shows the B-V color instead of the temperature. The color coding is the
following: red: 0 Gyr, green: 0.5 Gyr, blue: 1 Gyr, purple: 2 Gyr, light blue:
5 Gyr, yellow: 10 Gyr, black: 15 Gyr.}
\end{figure*}

Hence, these two types of models follow different aspects of the stellar
system. The first one describe the evolution of integrated properties
of the whole system. Therefore, these models are useful for unresolved
objects such as distant globular clusters in other galaxies.
The second describes the dynamical evolution of the stellar system.
Therefore, these models are useful for investigations of the dynamical
properties of a system, such as the velocity dispersion. \\

The ideal approach would be a combination of both a $N$-body simulation
and a stellar population synthesis model. In the case of globular clusters,
a good approach would be to assume that each simulation particle represents
one star. Since the stars have different masses, which may change due to
stellar evolution, the stars themselves may influence the dynamical properties
of the system. Effects such as mass segregation due to energy equipartition
may occur. This means that the high-mass stars sink to the center, whereas
the low-mass stars are more in the outer regions.
Conversely, the dynamical evolution may change the stellar
populations or even the stellar evolution. For example, a globular cluster
evolving in a tidal field of a host galaxy may remove in particular low-mass
stars from the cluster, which would cause a depletion of low-mass stars
leading to a bluer color of the cluster. \\

One should mention that there are other approaches. For example, Ivanova et al.
(2005) presented a Monte Carlo approach combining a population synthesis code
and a simple treatment of dynamical interactions in the dense cluster core
using a tool for three- and four-body interactions. This approach focusses
more on the evolution of binary fractions in globular clusters. This is in 
principle also possible with the approach mentioned above, but expensive in
terms of computation time. However, the first approach has the advantage that
the dynamical behavior of the whole cluster can be modeled more directly. \\

In this paper we follow the idea of combining $N$-body simulations with
stellar populations synthesis. The method is described in section
2. Results of test simulations are shown in section 3. 
In section 4 we compare these results to observational data of integrated
colors and spectra of globular clusters. Conclusions are drawn in section 5, 
and section 6 gives an outlook of our future work.

\section{Method}
As discussed above, we want to combine direct $N$-body simulations with stellar
population synthesis modeling in order to model the dynamical evolution
\em and \em color evolution of globular clusters at the same time. 
Following this approach, we are developing the HYDE-PARC 
(\underline{Hy}brid Code for \underline{D}ynamical \underline{E}volution 
and \underline{P}opulation \underline{A}nalysis of Stella\underline{R} 
\underline{C}lusters). \\

For this purpose the NBODY6++ code (Spurzem 1999) is used, which is a parallel
version of the Aarseth NBODY4 code (Aarseth 1999). Simple recipes to follow 
the changes of stellar masses, radii, and luminosities due to stellar 
evolution have been implemented into NBODY4 and NBODY6++ (Hurley et al. 2001), in the
sense that each simulation particle represents one star. These prescriptions 
cover all evolutionary phases and solar to globular cluster metallicities.
We used the stellar parameters obtained by this stellar evolution routine
and coupled them to the stellar library BaSeL 2.0 (Lejeune et al. 1997). \\

The $N$-body code uses a 
high-precision (4th order) direct $N$-body integrator, which allows us to follow 
stellar dynamical two-body relaxation as well as all kinds of close encounters 
between singles and binaries
 with high efficieny and precision and without any artificial change of the
 interaction potential (no softening). Such codes, containing a Hermite
scheme, hierarchically blocked individual variable time steps, Ahmad-Cohen
 neighbour scheme, and regularisations of two-body and higher encounters are
 being developed and supported by Aarseth (1985, 1999a, 1999b, for the general
 $N$-body code) and Mikkola (1997) and Mikkola \& Aarseth (1990, 1993, 1996,
1998, for the
 regularisation methods). We use the special version NBODY6++ (Spurzem 1999),
which is the only one available to use on massively parallel computers
(supercomputers  as well as PC Beowulf clusters, recently also including 
GRAPE accelerator cards). \\

The globular cluster is modeled as a spherical stellar system with a Plummer 
sphere. In future, rotating models will also be investigated, such as used
in Einsel \& Spurzem (1999).
The mass of each simulation particle (representing one ``star'') is
selected using a certain initial mass function (IMF). In the work presented here
a Kroupa et al. (1993) IMF is chosen. When the simulation is started, the stars
begin to evolve and follow their evolution as given by the stellar evolution
recipes (Hurley et al. 2000). In the language of stellar population
synthesis modeling this corresponds to single stellar evolution 
-- in future also binary evolution will be incorporated. 
While $N$-body
simulations are scale-free, stellar evolution requires us to fix the 
time scale. \\

The NBODY6++ code provides the stellar parameters mass $M$,  radius $r$ and 
bolometric luminosity $L$ of each star. By using 
\begin{equation}
\label{Teff}
  T_{\rm eff}^4 = \frac{L}{4 \pi r^2 \sigma} 
\end{equation}
and
\begin{equation}
\label{logg}
g = \frac{G M}{r^2}
\end{equation}
these parameters are transformed into $T_{\rm eff}$ and $\log g$. \\

\begin{figure*}
\centering
\includegraphics[width=17cm]{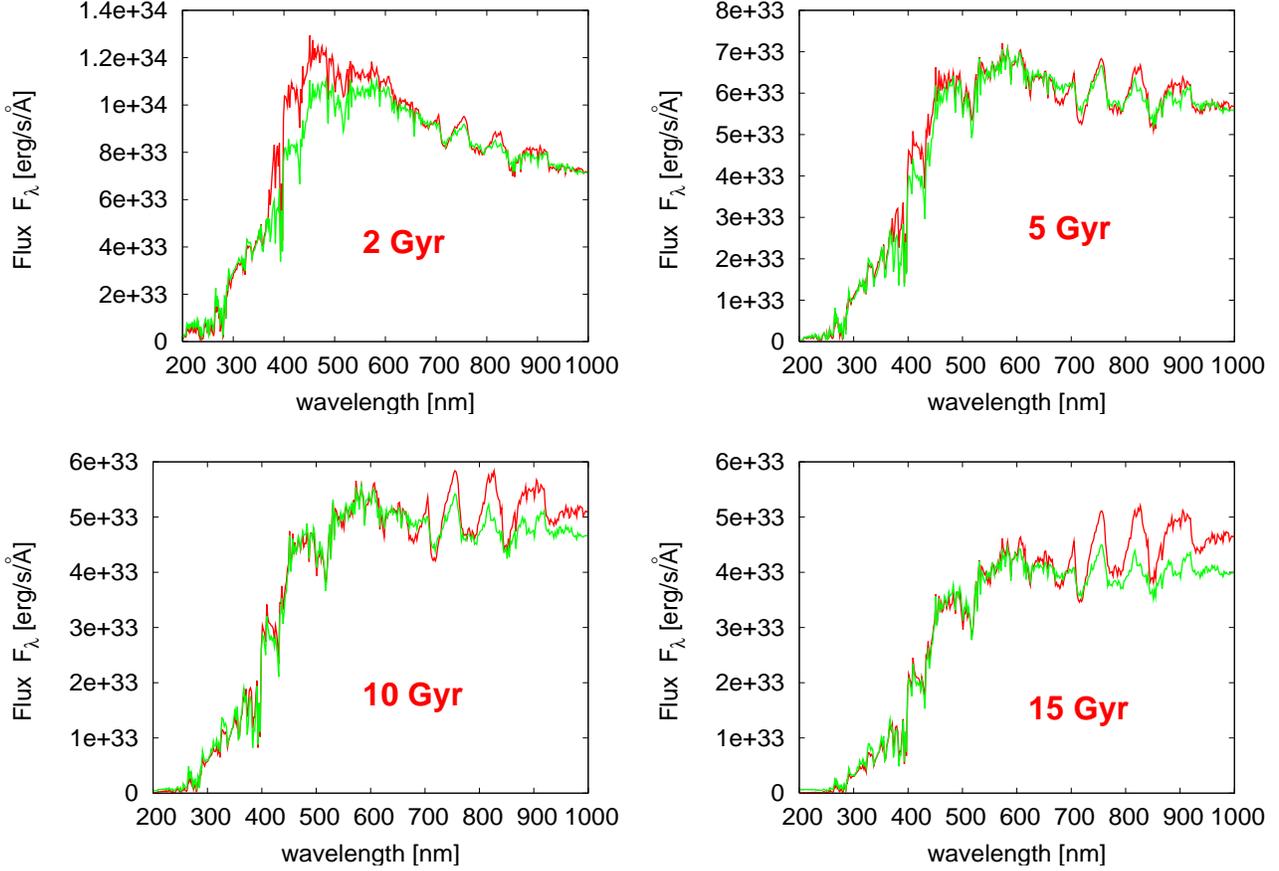}
\caption{Integrated spectrum of a isolated cluster at 4 different ages.
The red spectra show the result from the simulation of the NBODY6++ code.
The green spectra show the result from the PEGASE code.}
\end{figure*}

Since the NBODY6++ code deals with a single stellar population, the 
metallicity value, $Z$, is the same for all stars.  
Together with the effective temperature $T_{\rm eff}$ and the 
$\log g$ value delivered by the formula (\ref{Teff}) and (\ref{logg}) the 
simulation delivers a parameter triple for each star. \\

For a stellar library we use the BaSeL 2.0 library of Lejeune et al (1997).
It is organized as a grid defined in the 
$(Z, T_{\rm eff}, \log g)$ space. Some 
interpolation is necessary in order to
transform the library into a regular grid in terms of these parameters:
in particular for cool stars with temperatures below 3500 K we need to 
interpolate the surface gravity at some values in order to match the 
parameter values to the grid given by the higher temperature stars. 
In addition, at some metallicities there are some ``holes'' in the parameter
grid, and therefore we need to interpolate in order to fill the mesh
point gaps. \\ 

Hence, the library grid fits to the parameter triple described above. 
Therefore, this  parameter triple can be interpolated to the parameter 
grid of the library. This leads to a spectrum that can be attached to each 
star. \\

For example, Figure 1 shows the CMD for a simulation 
of 50000 particles. The effective temperature plotted in the left diagram is 
converted into the B-V color in the right diagram by using the BaSeL 2.0 
library. \\

The spectra of all stars can be added in order to get the integrated spectrum
of the entire stellar population. A convolution of the integrated spectrum with
filter curves delivers integrated colors. For test purposes so far only the
Johnson filter set is used. 

\section{Results of test simulations}
For test purposes we investigate several simulations. For the mass distribution
we use a Kroupa IMF (Kroupa et al. 1993) with an upper mass of 8.16 
M$_\odot$ and a lower mass of 0.16 M$_\odot$. We use only single stars 
without primordial binaries. Also no binary evolution is taken into account.
We use a Plummer model, both for the
isolated cluster and also for the cluster in the tidal field. In the latter
case, for the external force a standard tidal field is used, that corresponds
to a cluster orbiting at a distance of 8.5 kpc around the galactic center in a 
circular orbit. For most cases we use initially 50000 particles, except
in cases where other particle numbers are mentioned.

\begin{figure*}
\centering
\includegraphics[width=17cm]{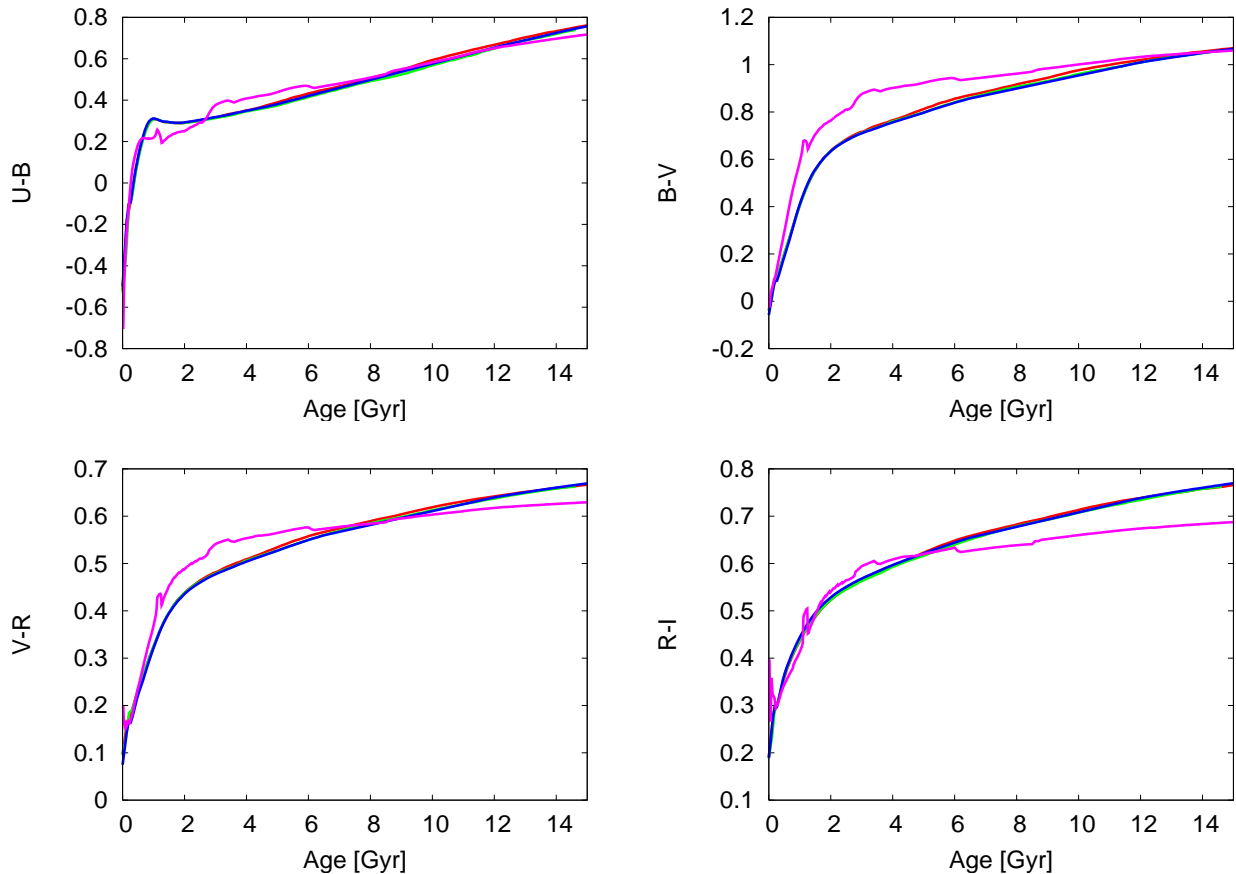}
\caption{Integrated colors of a isolated cluster as a function of age.
Color coded are different particle numbers: red: 10000 particles, 
green: 20000 particles, blue: 50000 particles. As a comparison, the purple
lines show the time evolution of colors calculated with the PEGASE code.}
\end{figure*}

\subsection{Comparison with the PEGASE code}
First we will concentrate on the integrated spectra and colors
of an isolated cluster. Since the BaSeL 2.0 library is also used in the 
PEGASE code (Fioc \& Rocca-Volmerange 1997) it is also interesting to compare
the results of the NBODY6++ simulations to the results from the PEGASE code.
Figure 2 shows the integrated spectra at the ages 2 Gyr, 5 Gyr, 10 Gyr and
15 Gyr. At low ages there is a difference between the NBODY6++ result and
the PEGASE result in the regime around 400 nm - 600 nm. This difference 
vanishes towards higher ages. In the wavelength regime above 700 nm there 
is a mismatch that increases toward higher ages. \\

The time evolution of integrated colors shown in Figure 3 reflects 
consequently  the mismatches mentioned above. The difference in B-V color
reaches its maximum at around 3 Gyr. Towards higher ages the NBODY6++ result
and the PEGASE result are converging. Corresponding to the behavior at higher
wavelengths the R-I colors of both codes match best at lower ages.
Towards higher ages there is an increasing mismatch. As seen in Figure 2, in 
the lower wavelength regime the results of both codes are in good agreement. 
This is reflected in the time evolution of the U-B color in Figure 3, which
shows good agreement in the results of both codes at all ages. \\

The difference between the PEGASE results and the NBODY6++ results is most likely 
due to the different treatment of stellar evolution. One reason is the issue 
of the thermally-pulsing asymptotic giant branch (TP-AGB) stars. 
These stars are rare, but bright, and therefore already a 
few of these stars in the whole cluster are enough to influence the integrated
spectrum, in particular at longer wavelengths. In the current version 
presented here the TP-AGB stars are not included. This issue will be addressed
in the future work. \\

As expected for an isolated cluster, the time behavior of all colors
is practically independent of the particle number. \\

The results discussed do far are obtained at solar metallicity. The metallicity
dependence of integrated colors is shown in the following figures.
Figure 4 shows the metallicity dependence of different optical colors at 
different ages. A comparison of the results of both codes still shows good 
agreement at most metallicities. For solar metallicities it reflects the 
result seen in Figure 3: the agreement is better for higher ages. 
Interestingly, there is a certain mismatch for lower metallicities.
This low metallicity mismatch is even higher for V-R colors, and also present
for R-I colors. \\

\begin{figure*}
\centering
\label{metall_U_B}
\includegraphics[width=17cm]{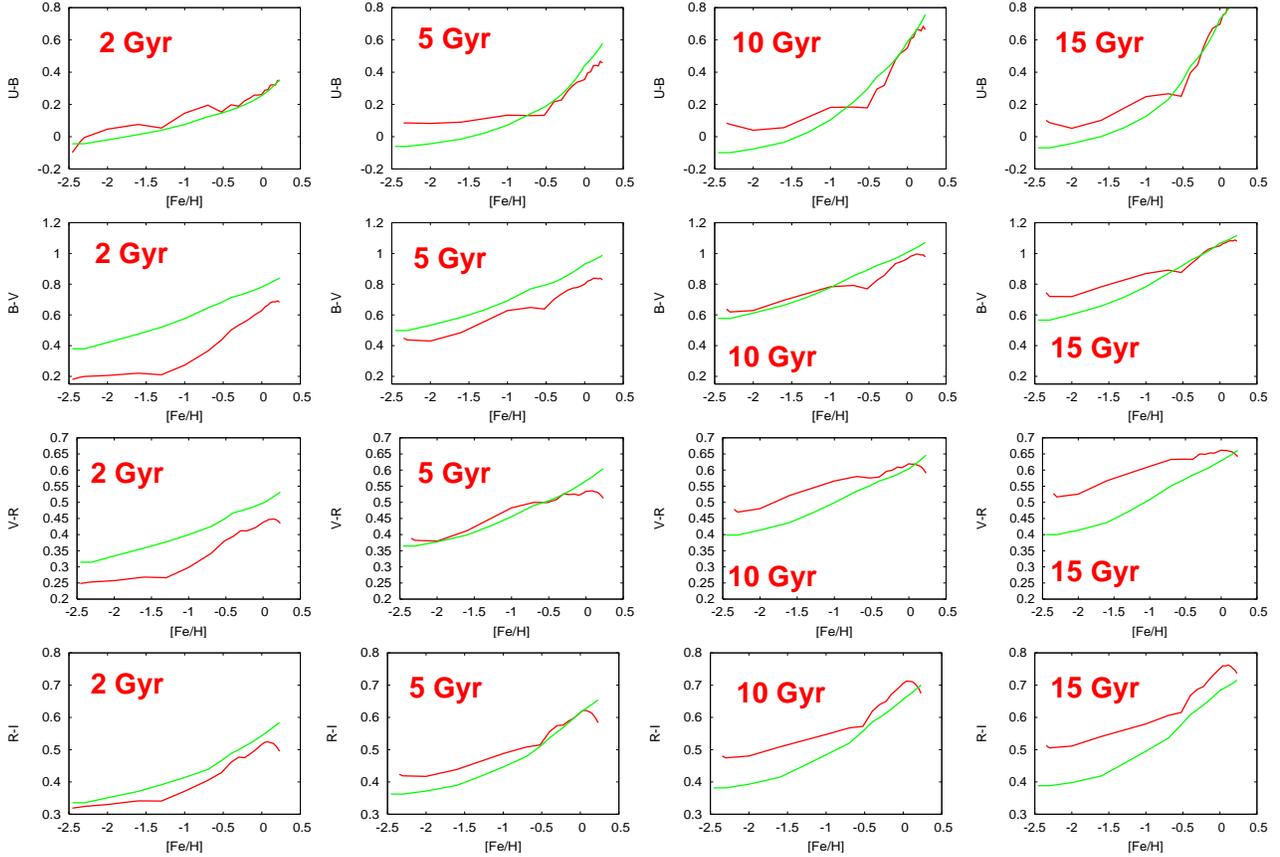}
\caption{Integrated colors of a isolated cluster as a function of metallicity.
Each row shows one color according to the colors: U-V, B-V, V-R, R-I.
Each column shows one age of the cluster.
The color coding means the following: red: NBODY, 
green: PEGASE result.}
\end{figure*}

\subsection{Influence of the tidal field}
 We also investigated the influence of the tidal field of a host galaxy on
the integrated color evolution of the globular cluster. 
Owing to the tidal
field particles are removed from the cluster when leaving the tidal radius.
Therefore the cluster is disrupted and the particle number is monotonically
decreasing. The result of the
integrated B-V color is shown in Figures 5 and 6. Figure 5 shows the results
for different particle numbers. As a result, the color evolution of a
tidally disrupted cluster follows more or less the evolution of a isolated 
cluster for most of its lifetime. Only shortly before the cluster will be 
dissolved, it becomes first slightly bluer and then it becomes dramatically
redder. \\

Figure 6 shows the result for different $\bar{R}$ values. The $\bar{R}$ 
value gives the virial cluster radius in pc. Therefore it controls
the concentration of the globular cluster by a given distance of the cluster
center to the gravity center. Note, that a variation of $\bar{R}$ is
physically equivalent to a variation of the galactocentric distance.
However, this is only true in the dynamical sense in a scale-free language.
But in the simulation here the stellar evolution sets a second timescale,
and therefore the physical timescale is fixed and the system can no longer
be considered as being scale-free. Since the $\bar{R}$ value influences the
dynamical timescale, whereas the stellar evolution timescale is fixed, 
a certain physical time of the system mean different NBODY6++ time units in
simulations with different $\bar{R}$ values. \\

\begin{figure*}
\centering
\includegraphics[width=8cm]{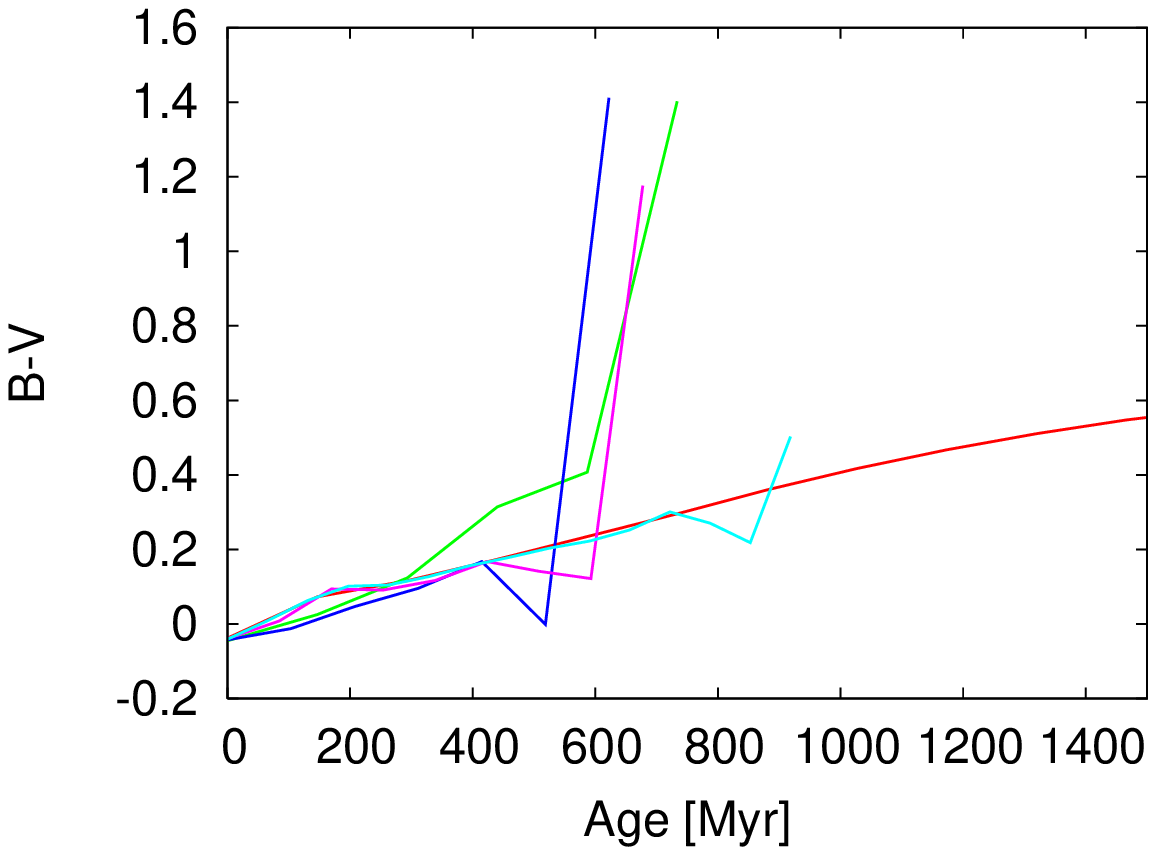}
\includegraphics[width=8cm]{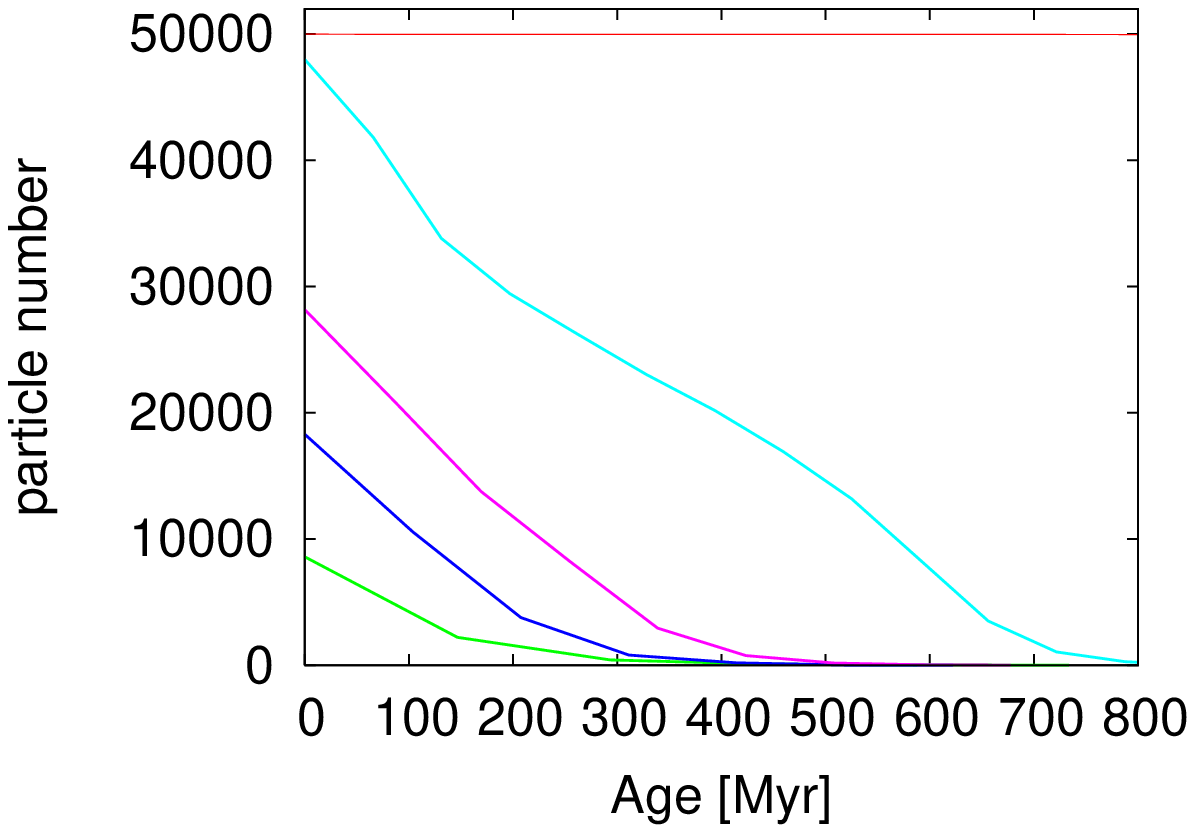}
\caption{Color evolution of the globular cluster influenced by the tidal
field of the host galaxy. This simulation is done for a virial cluster
radius of $\bar{R} = 30$ pc.
Color coded are different initial particle numbers.
Left plot: B-V color, right plot: time evolution of the particle number.
Color coded are different initial particle numbers: green: 10000, blue: 20000,
purple: 30000, light blue: 50000. As a comparison, the red line shows a 
isolated cluster. }
\end{figure*}

\begin{figure*}
\centering
\includegraphics[width=8cm]{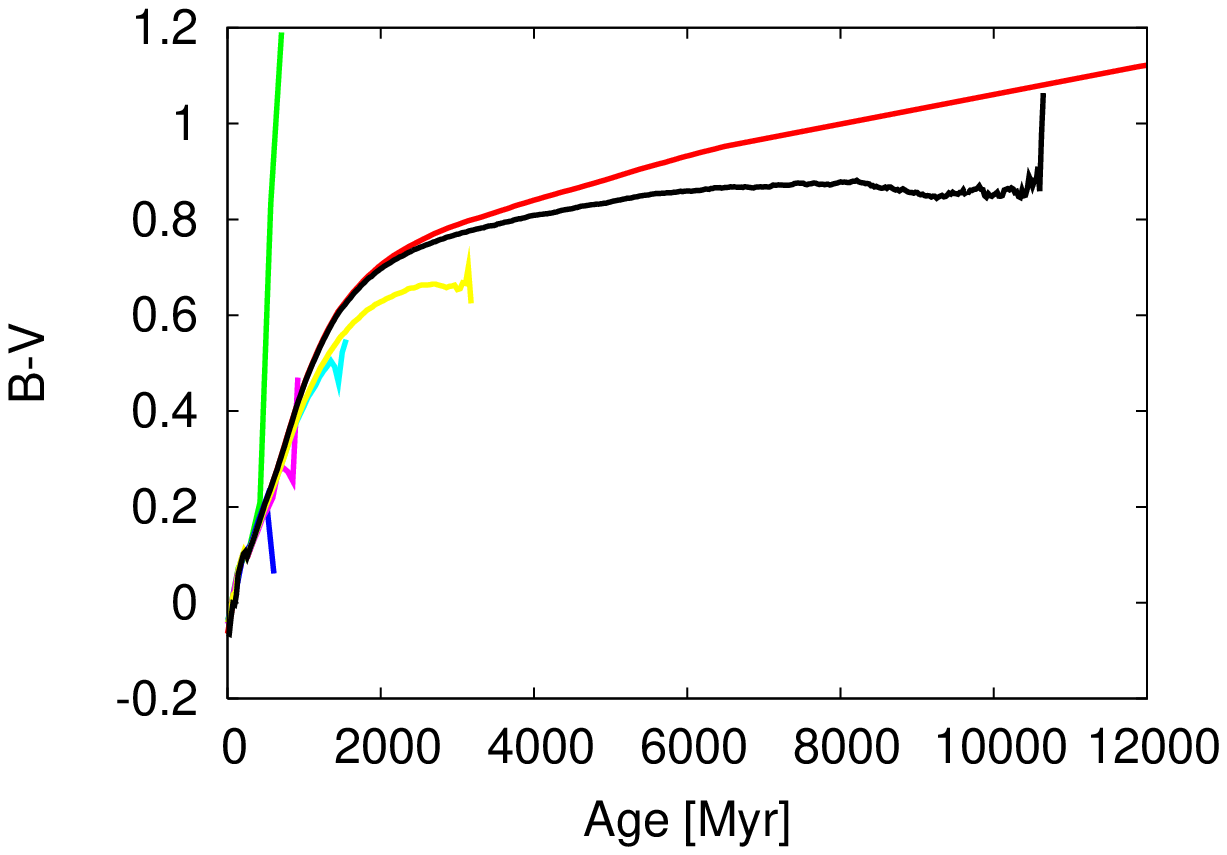}
\includegraphics[width=8cm]{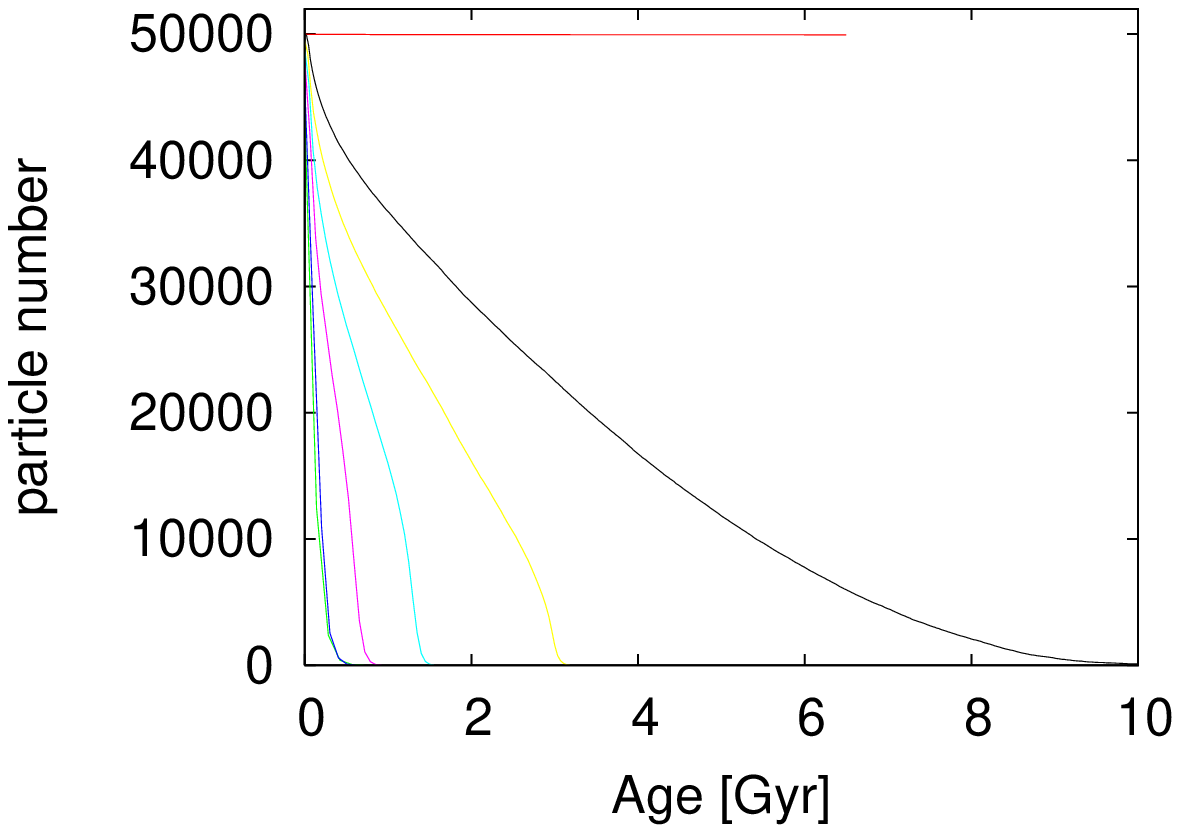}
\caption {Color evolution of the globular cluster influenced by the tidal
field of the host galaxy.
Left plot: B-V color, right plot: time evolution of the particle number.
Color coded are different virial cluster radii, given in the code as 
$\bar{R}$ values: green: $\bar{R}$ = 50, 
blue: $\bar{R}$ = 40, purple: $\bar{R}$ = 30, light blue: $\bar{R}$ = 25, 
yellow: $\bar{R}$ = 20, black: $\bar{R}$ = 15.
As a comparison, the red line shows a isolated cluster.}
\end{figure*}

For large $\bar{R}$ values corresponding to a low
cluster concentration the evolution shows a similar behavior as seen in Figure
5. However, when the $\bar{R}$ value is low enough, the cluster lifetime 
increases drastically, and becomes significantly bluer compared to the isolated
cluster. This is due to the effect of mass segregation. In order to investigate
this effect, Figure 16 shows the average particle number for both simulations,
for an isolated cluster and for the simulation with the lowest $\bar{R}$
value. The isolated cluster shows a decreasing average mass due to stellar
evolution effects. The cluster in the tidal field shows a strong upturn in
average mass. This is due to the effect of mass segregation. Resulting from
this the low-mass stars escape from the cluster, and therefore the average
mass increases up to 1 solar mass. Hence, the origin for the bluer color
is a loss of low-mass particles in the stellar population. \\

\begin{figure*}
\centering
\includegraphics[width=8cm]{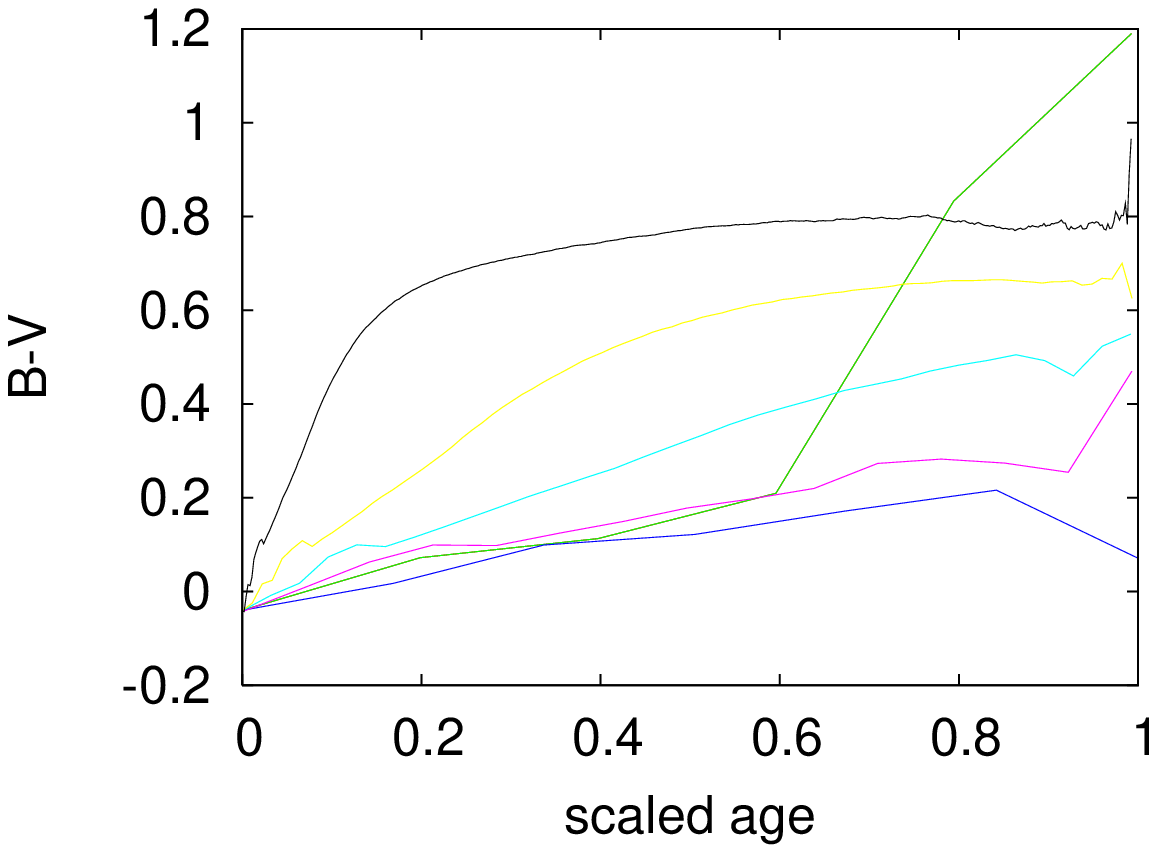}
\includegraphics[width=8cm]{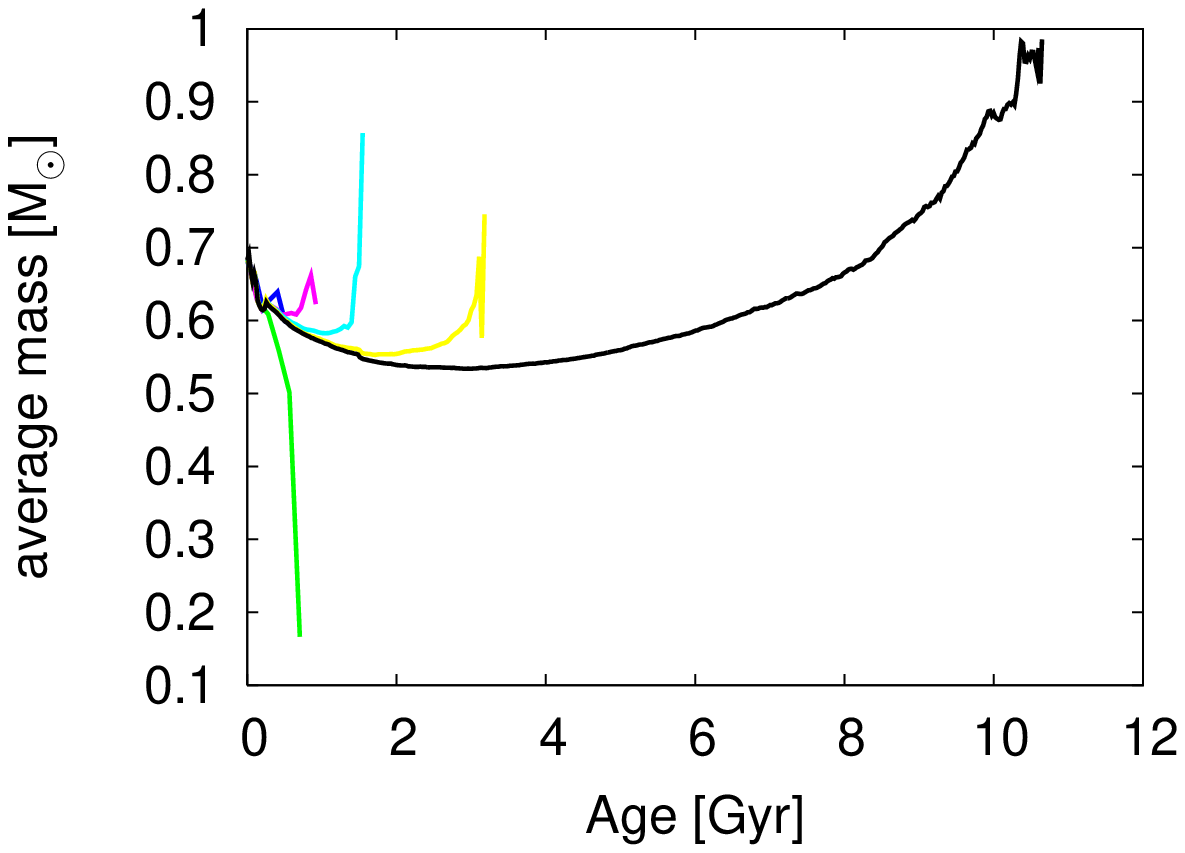}
\caption{Left plot: Same as Figure 6 (left). This time the age is scaled to 
the dissolution time. 
Therefore, for every simulation at age = 1 the cluster is dissolving.
Color coded are different  $\bar{R}$ values: green: $\bar{R}$ = 50, 
blue: $\bar{R}$ = 40, purple: $\bar{R}$ = 30, light blue: $\bar{R}$ = 25, 
yellow: $\bar{R}$ = 20, black: $\bar{R}$ = 15. Right plot: evolution of average
mass per particle. The color coding is the same.}
\end{figure*}

\begin{figure*}
\centering
\includegraphics[width=16cm]{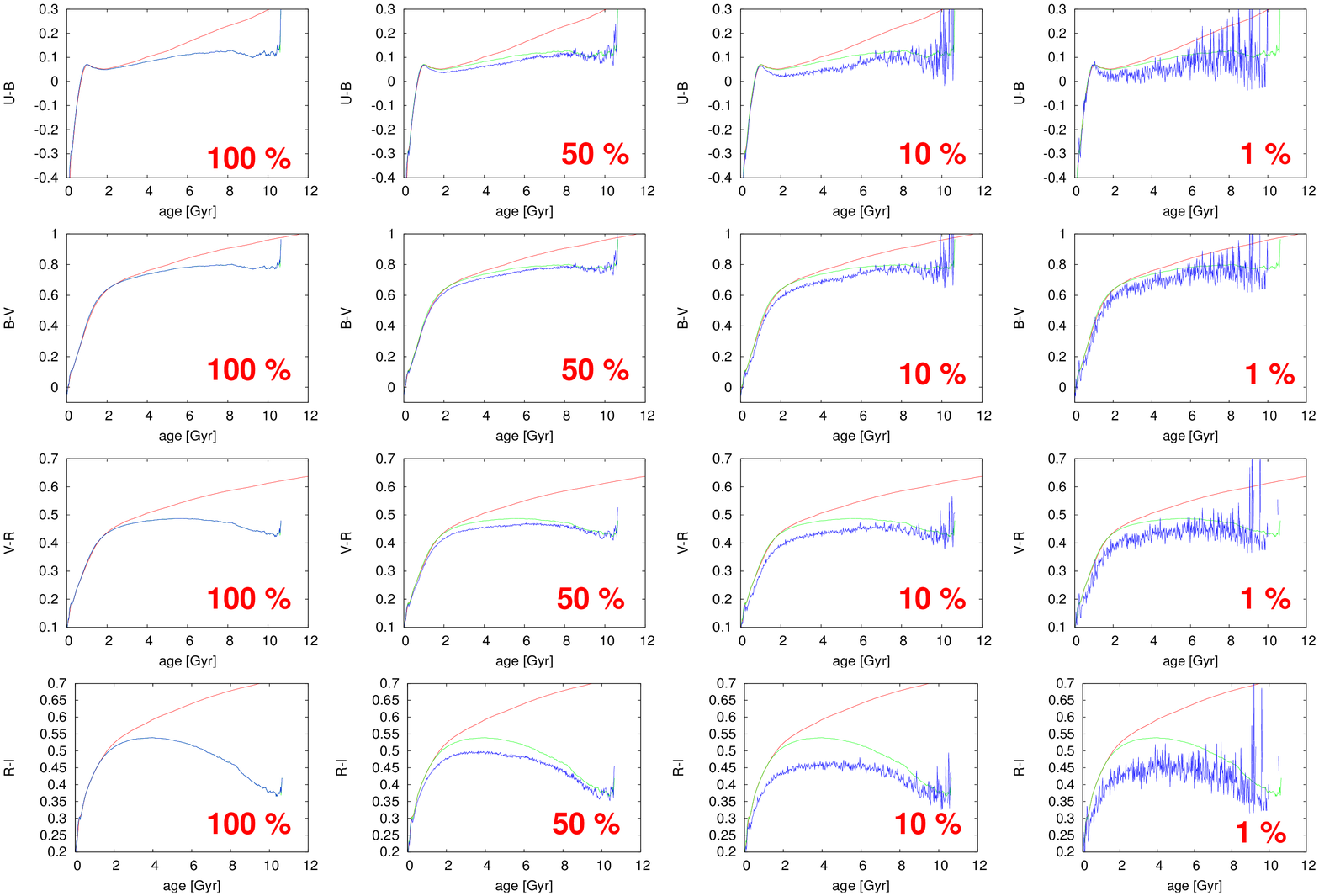}
\caption{Colors of the globular cluster influenced by a tidal field at
different Lagrange radii. The green curves include all particles of the whole
cluster. The blue curves include all particles within a certain Lagrange
radius. Each row shows one color according to the colors: U-V, B-V, V-R, R-I.
The four  columns correspond to four different Lagrange radii: 100\%,
50 \%, 10\% and 1\%. As a comparison, the red curve shows the color evolution 
of the isolated cluster.}
\end{figure*}

At the time of dissolution the particle number naturally becomes very low. Due to 
statistical fluctuations the average particle mass may become very low or
very high. This is shown in Figure 7 (right), where the average particle mass
is shown for the same set of $\bar{R}$ values as shown in Figure 6
For example, at $\bar{R} = 50$ there are only two particles in the last time
before the simulation ended, and both are low mass stars, resulting in a low
average particle mass. This causes a strong jump in B-V color, as shown in
Figure 6. On the other hand, for $\bar{R} = 25$ there are 9 particles
at the final time. Among them, there are two particles with masses around
0.9 solar masses, resulting in a higher average mass at the last time step. \\

We investigated the dependence of the mass-to-light ratio on the B-V color.
It is shown in Figure 15. The left plot shows a simulation of a globular 
cluster with 50000 particles. The result can be well described as a straight
line fitted to the data. The right plot shows a simulation of a globular
cluster with $\bar{R}$=15 influenced by a tidal field. From Figure 6 we know 
that
this cluster also lives for several Gyrs and shows bluer colors compared
to the isolated cluster. In Figure 15 this effect of bluer color corresponds to
a sharp edge around B-V = 0.7. At this edge there is a strong deviation
in M/L ratio, when comparing to the color-M/L correlation of the isolated
cluster. The reason for this M/L ratio deviation is a strong increase in the
average mass at that color. This is shown in Figure 16 (right), that exactly at
this color the average mass increases due to the effect of the escape of
low-mass stars in clusters in a tidal field. Since these low-mass stars have
a higher M/L ratio, the cluster M/L ratio is decreasing, when the cluster
loses low-mass stars. \\

\begin{figure*}
\centering
\includegraphics[width=16cm]{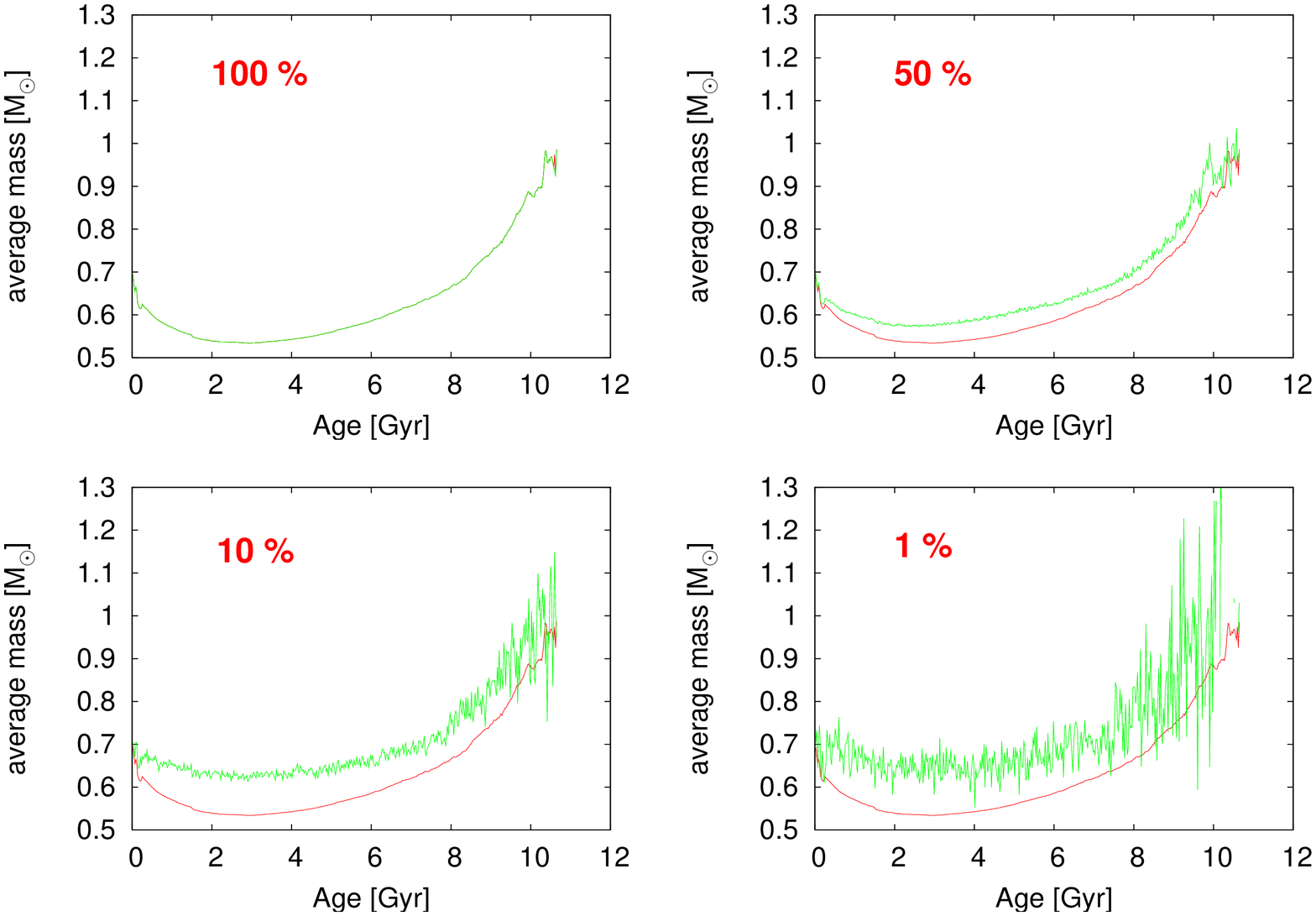}
\caption{Time evolution of the average mass per star of the globular cluster 
influenced by a tidal field at different Lagrange radii. The red curves 
include all particles of the whole cluster. The green curves include all 
particles within a certain Lagrange radius. The four plots correspond to four 
different Lagrange radii: 100\%, 50 \%, 10\% and 1\%.}
\end{figure*}

\begin{figure*}
\centering
\includegraphics[width=16cm]{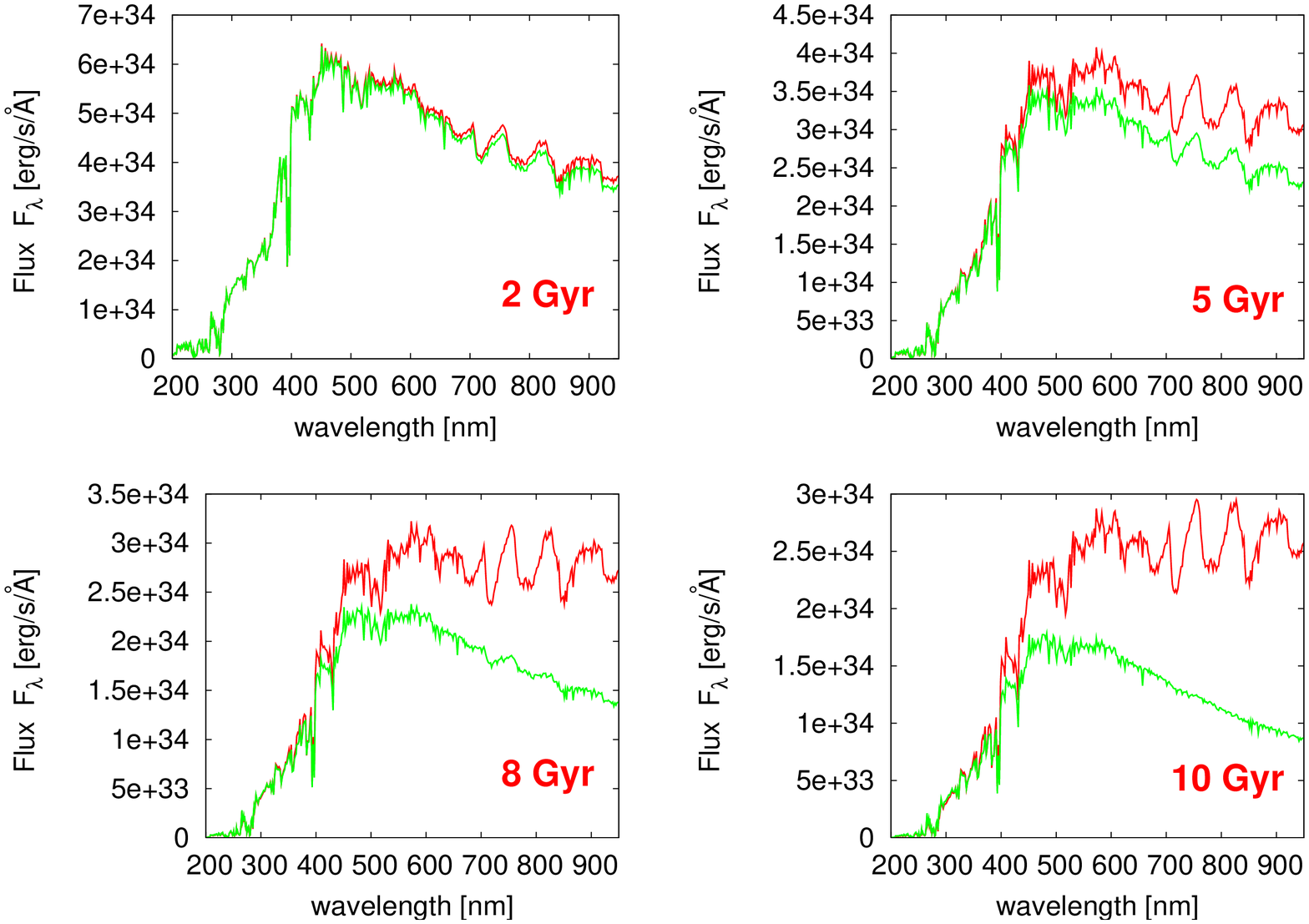}
\caption{Integrated spectra of stellar clusters at different time steps.
The red spectra show an isolated cluster, whereas the green spectra show 
a cluster in a tidal field.}
\end{figure*}

It is interesting to notice that Figure 16 (right) shows that the color remains
constant during the period when the M/L ratio is increasing. This reflects
a balance between two different effects. Firstly, the cluster becomes redder
due to the effect of stellar evolution and, secondly, the cluster becomes bluer
due to the loss of red low-mass stars. Since the total luminosity is dominated 
by the main sequence turnoff, it is able to balance the loss of low-mass stars,
which are larger in number, but also fainter. Altogether, it turns out that
in this phase the integrated color of the cluster is close to constant. This
effect is already seen in Figure 6. The black curve in this figure shows
little evolution in B-V color at ages from 3 to 10 Gyr. \\

The results presented so far include all particles contained in the 
simulated cluster. In Figure 8 we show the dependence of color evolution 
on radial position from the cluster centre, by utilising the Lagrangian radii. 
The Lagrange radius gives us at each timestep the radius containing
a certain fraction of the mass of the cluster. Four different Lagrange radii
are shown containing 100 \% of the cluster mass down to 1 \% of the cluster 
mass. The plots show the same simulation as discussed in Figures 6 and 15,
which is a cluster in a tidal field with $\bar{R}$ = 15. As shown in Figure 8,
 the difference is negligible for U-B colors and for B-V colors,
whereas there is a strong radius dependence in the R-I colors. In 
Figure 9 we investigated the average stellar mass at 
different Lagrange radii. As a result, in the inner 10\% there is a parallel
shift in the sense that the average stellar mass is higher in the inner regions.
Hence, the higher mass stars are more located in the inner regions, whereas
the lower mass stars are more located in the outer regions. This effect 
reflects
the decreasing R-I color in the inner 10\%: In the longer wavelength regime
we expect to see the red, low mass, stars and therefore the R-I color becomes
bluer in the inner regions, where these stars become underrepresented compared
to the higher mass stars.  \\

So far we have only looked at the integrated colors. It is also interesting
to look at the spectra. Figure 10 compares the integrated spectrum of an
isolated stellar cluster with the spectrum of a cluster in a tidal field
with $\bar{R} = 15.$ As we have seen in Figure 2, the isolated cluster behaves
like a simple stellar population modeled by usual stellar population methods.
In comparison, the cluster in the tidal field loses more and more flux
at longer wavelengths. This is what we already noticed in the color plots
-- the cluster becomes bluer compared to the isolated cluster.
In addition, the spectral features at longer wavelengths become less 
pronounced at higher ages, compared to the isolated cluster. \\

Figure 11 shows the spectra at different Lagrange radii. Here we see again
the effect of mass segregation, that alters the spectra in the inner part
of the cluster in the sense that the fluxes at higher wavelengths are lowered
in the inner part of the spectrum. This is due to the lack of red low-mass
stars in the cluster center. \\

So far, all simulations are done with the same metallicity of [Fe/H] = 0
and with the same galactocentric distance, in the sense that the cluster is
orbiting at the position of the Sun with respect to the Galactic center.  
Figure 12 shows simulations of clusters
with different metallicities. The influence of the galactocentric distance is
shown in Figure 13. It shows simulations of clusters circulating around
the center at a distance of 4 kpc and at 8.5 kpc. At 5 Gyr the cluster with
a radius of 4 kpc is dissolved, whereas the cluster at 8.5 kpc lives for
about 10 Gyr until the cluster is dissolved. One can see from Fig.
12 and Fig. 13 that both effects have a dramatic effect on the spectra, in
particular at higher ages.  \\

We also investigate the radial distribution of integrated colors. Figure 14
shows these color distributions. Since we deal with relatively low particle 
numbers, which
become even lower at higher ages, we decided to plot cumulative radial
distributions. We show both the radial distribution in 3D space, and the
projected radial distribution in the 2D plane as well. Whereas the 3D
distribution has a more physical meaning, the projected 2D distribution
brings our results closer to the observations, because clusters can only
be observed as a 2D projection on the sky. As we can see in these plots,
the 2D projected profiles are less steep than the 3D profiles. This is of
course a projection effect. \\

In Figure 24 we show cumulative plots of the radial distribution of different
stellar types. Whereas the total number of main sequence stars is slightly
decreasing, the number of red giants increases by a factor of 3 between 2 Gyr
and 5 Gyr. The number of AGB stars is low, and the number of white dwarfs
increases monotonically. Figure 25 shows the same cumulative plots but with 
the number counts are normalized to 1. \\

\begin{table}
\caption{Numbers of different stellar types in a simulation of a 
globular cluster in a tidal field with a initial particle number of 50000. 
At 4 different ages the total number of stars of a certain type is shown, 
and the number of stars, which
are just escaping at that time. The different stellar types are:
1 = main sequence stars, 2 = red giants, 3 = AGB stars , 4 = white dwarfs, 
5 = neutron stars, 6 = black holes.}
\centering
\begin{tabular}{c c c c c c c c c}
\hline
Star & \multicolumn{2}{c}{1 Gyr} &  \multicolumn{2}{c}{2 Gyr} 
& \multicolumn{2}{c}{5 Gyr} &  \multicolumn{2}{c}{8 Gyr} \\
type & total & esc.  & total & esc. & total & esc.  & total & esc. \\
\hline 
 1 & 34240 &   248  &   26287 &  234 &  9485 & 108 & 1357 & 33 \\
 2 &    28 &     0  &     58  &   1  &    77 &  0  &  40  &  1 \\
 3 &     9 &    1   &      8  &   0  &     1 &  0  &   1  &  0 \\
 4 &  1687 &   17   &    2427 &   32 &  2218 & 46  & 670  & 15 \\
 5 &     0 &    0   &      0  &   0  &     0 &  0  &   0  &  0 \\
 6 &     0 &    0   &      0  &    0 &     0 &  0  &   0  &  0 \\
\hline
\end{tabular}
\end{table}

\begin{figure*}
\centering
\includegraphics[width=16cm]{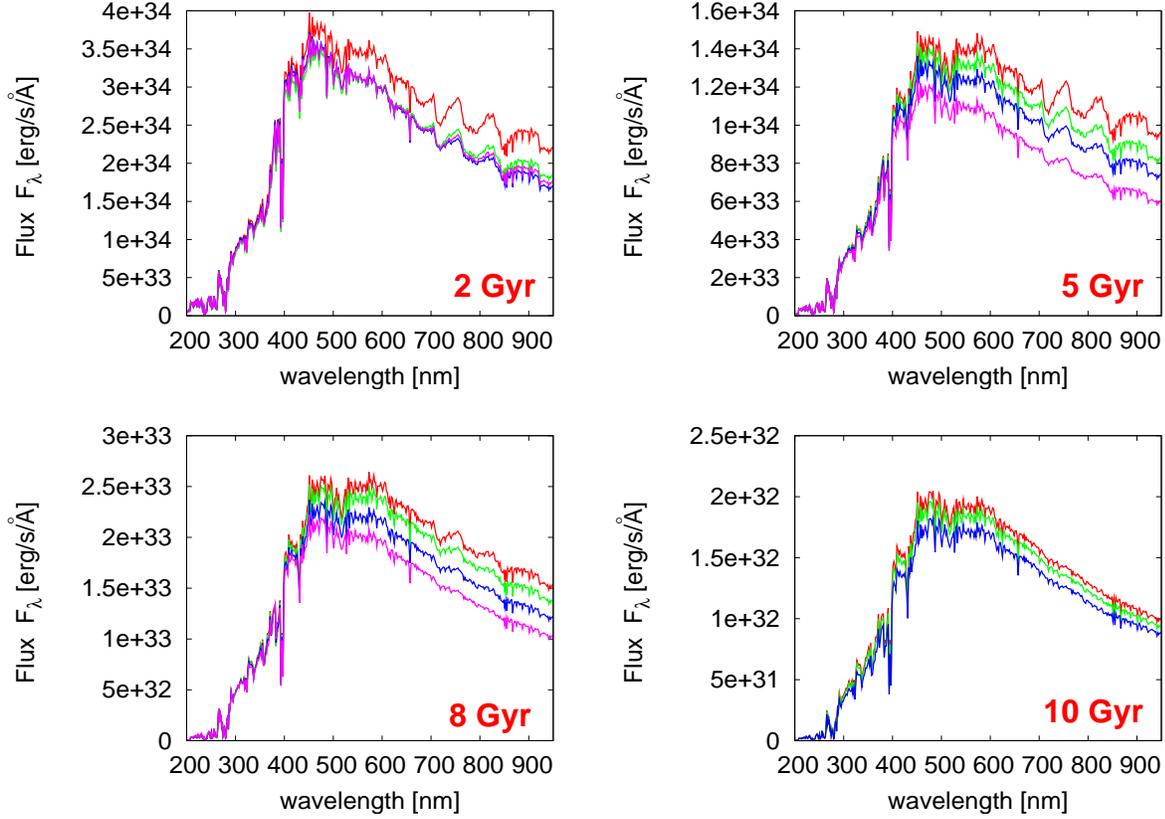}
\caption{Spectra of the cluster in the tidal field, as shown in the green 
spectrum in Figure 10. Color coded are different Lagrange radii. Red: 100\%
enclosed mass, green: 50\%, blue: 10\%, purple: 1\%.}
\end{figure*}

\begin{figure*}
\centering
\includegraphics[width=16cm]{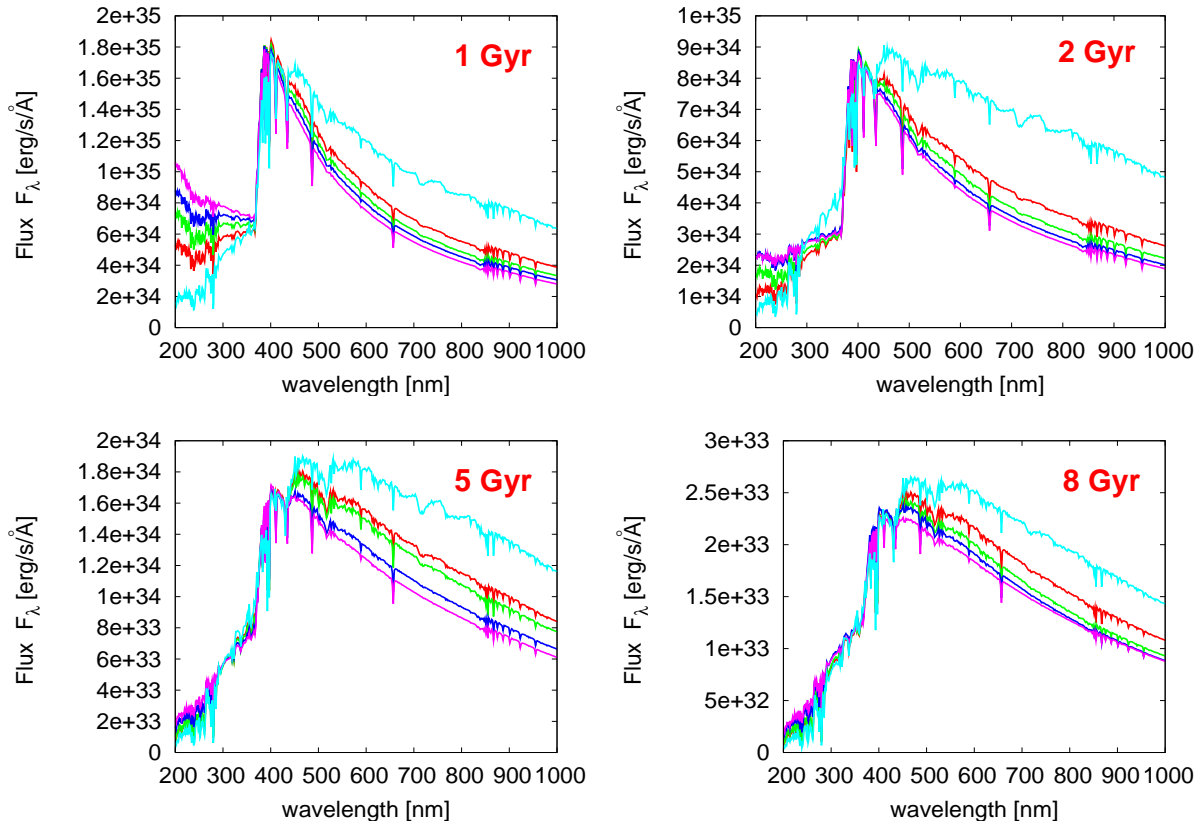}
\caption{Integrated spectra of a clusters in a tidal field with different
metallicities. The color coding is as following: red: [Fe/H] = -0.7, 
green: [Fe/H] = -1.0, blue: [Fe/H] = -1.2, purple: [Fe/H] = -1.5, light blue: [Fe/H] = 0.0.}
\end{figure*}

\begin{figure*}
\centering
\includegraphics[width=16cm]{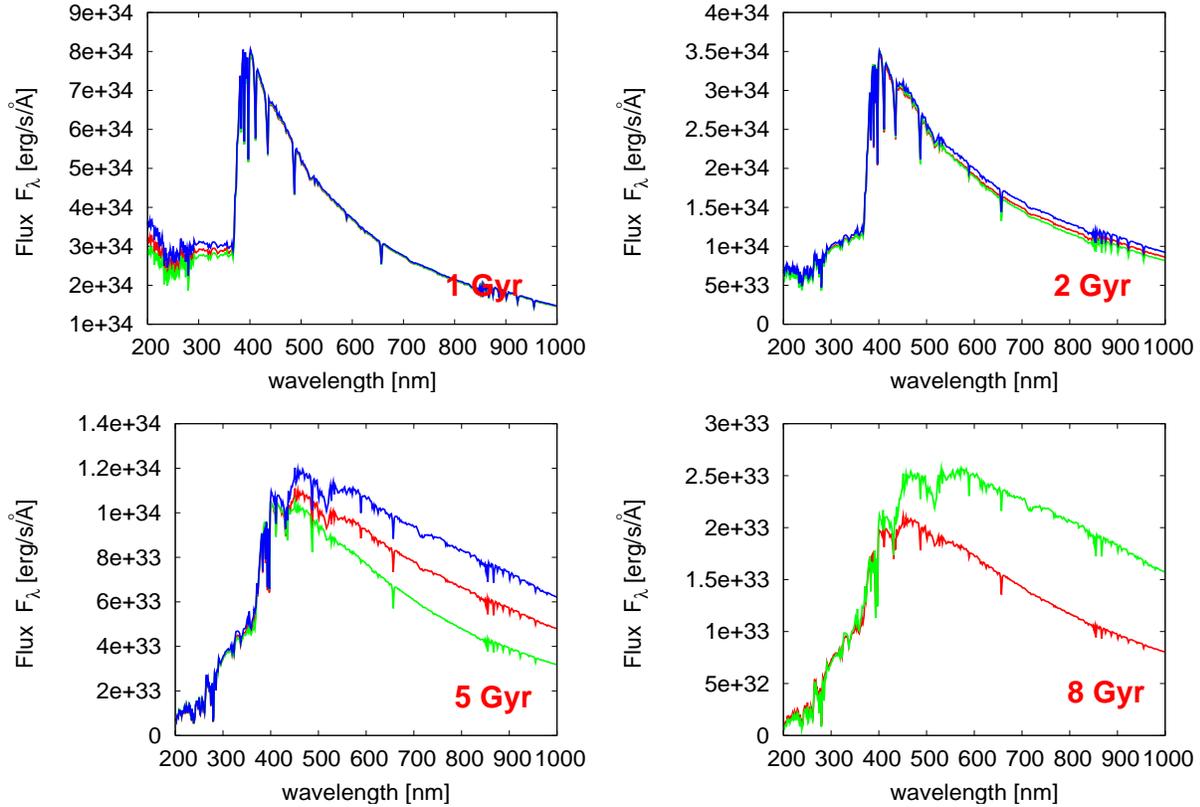}
\caption{Integrated spectra of a clusters in a tidal field with diffferent
galactocentric distances. The color coding is as following: red: R = 8.5 kpc, 
green: R = 4 kpc, blue: R=16 kpc. In order to arrange the cluster lifetime
in a useful range, the $\bar{R}$ values are chosen in a certain way:
 $\bar{R}$ = 15 for R = 8.5 kpc, $\bar{R}$ = 10 for R = 4 kpc, and
$\bar{R}$ = 20 for R = 16 kpc. }
\end{figure*}

\begin{figure*}
\centering
\includegraphics[width=16cm]{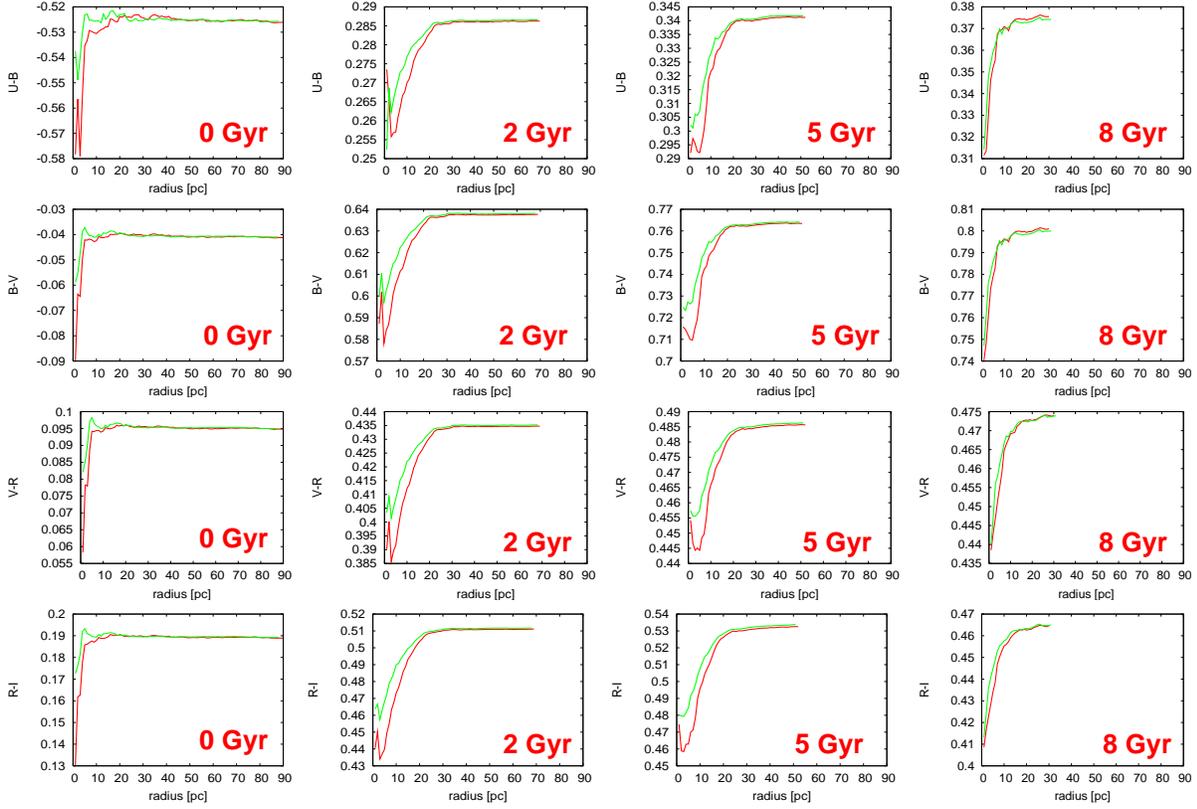}
\caption{Cumulative color radial distribution of the cluster in the tidal
field. Each row shows one color according to the colors: U-V, B-V, V-R, R-I.
Each column shows one age of the cluster.
The red curve shows the color distribution of all particles which are 
enclosed in concentric spheres of different radii. The green curve shows
the color distribution which are enclosed in concentric circles in the
XY-plane.}
\end{figure*}

\begin{figure*}
\centering
\includegraphics[width=8cm]{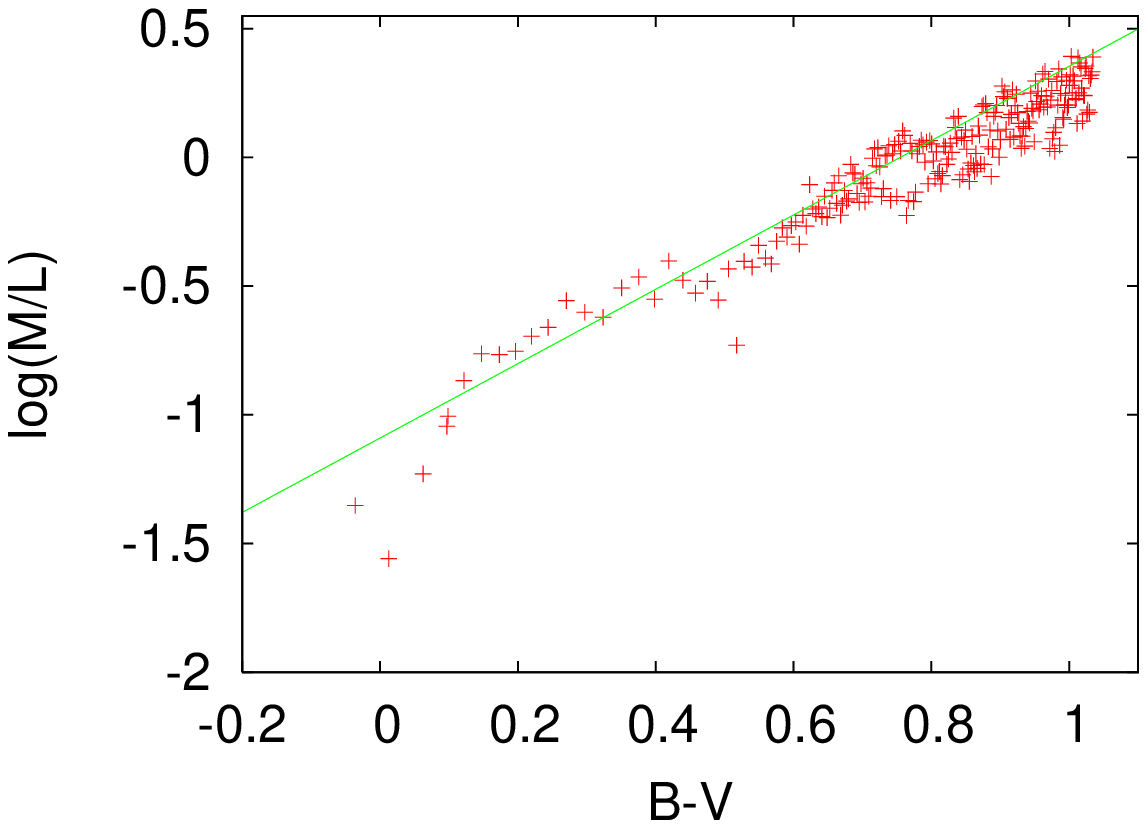}
\includegraphics[width=8cm]{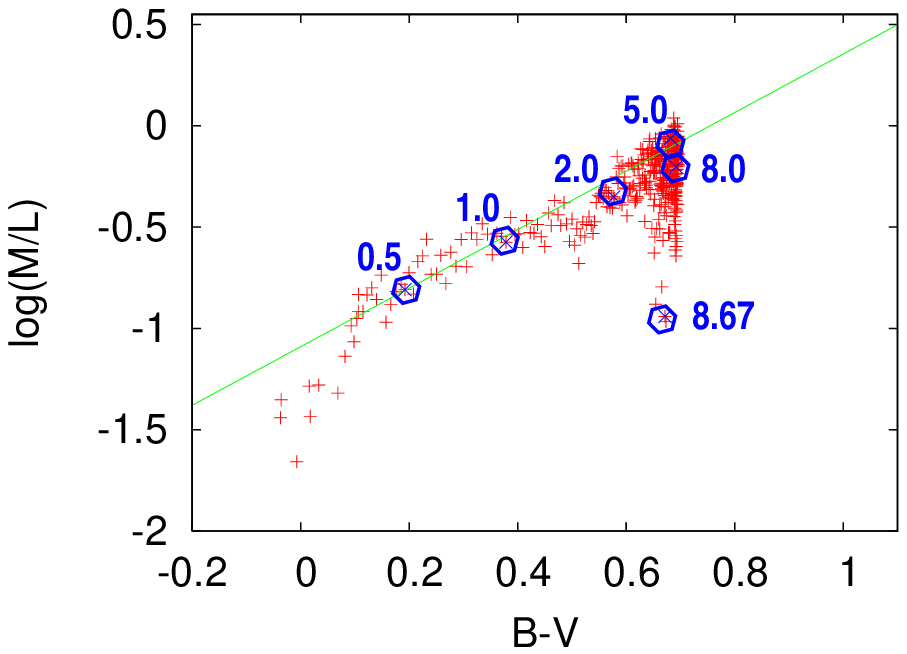}
\caption{B-V colors versus mass to light ratio for a globular cluster with
50000 particles. Each dot represents the integrated B-V color and the total
M-L value for a certain timestep. The left plot shows a isolated cluster.
The right plot shows a cluster influenced by a tidal field with $\bar{R}$ = 15,
wich corresponds to the black line in Figure 6. The green line in both
plots is a least square fit to the points in the left plot. The blue numbers
in the right plot label certain ages in Gyr.}
\end{figure*}

\begin{figure*}
\centering
\includegraphics[width=8cm]{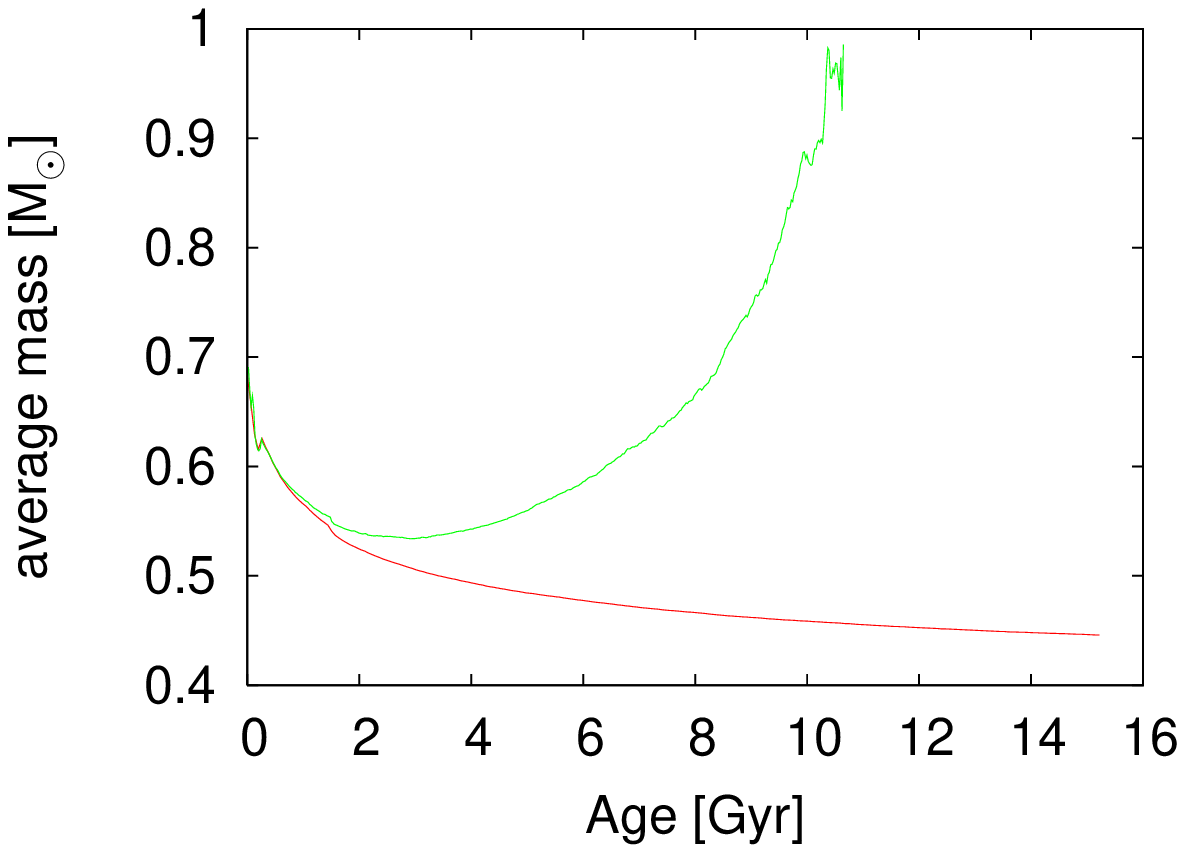}
\includegraphics[width=8cm]{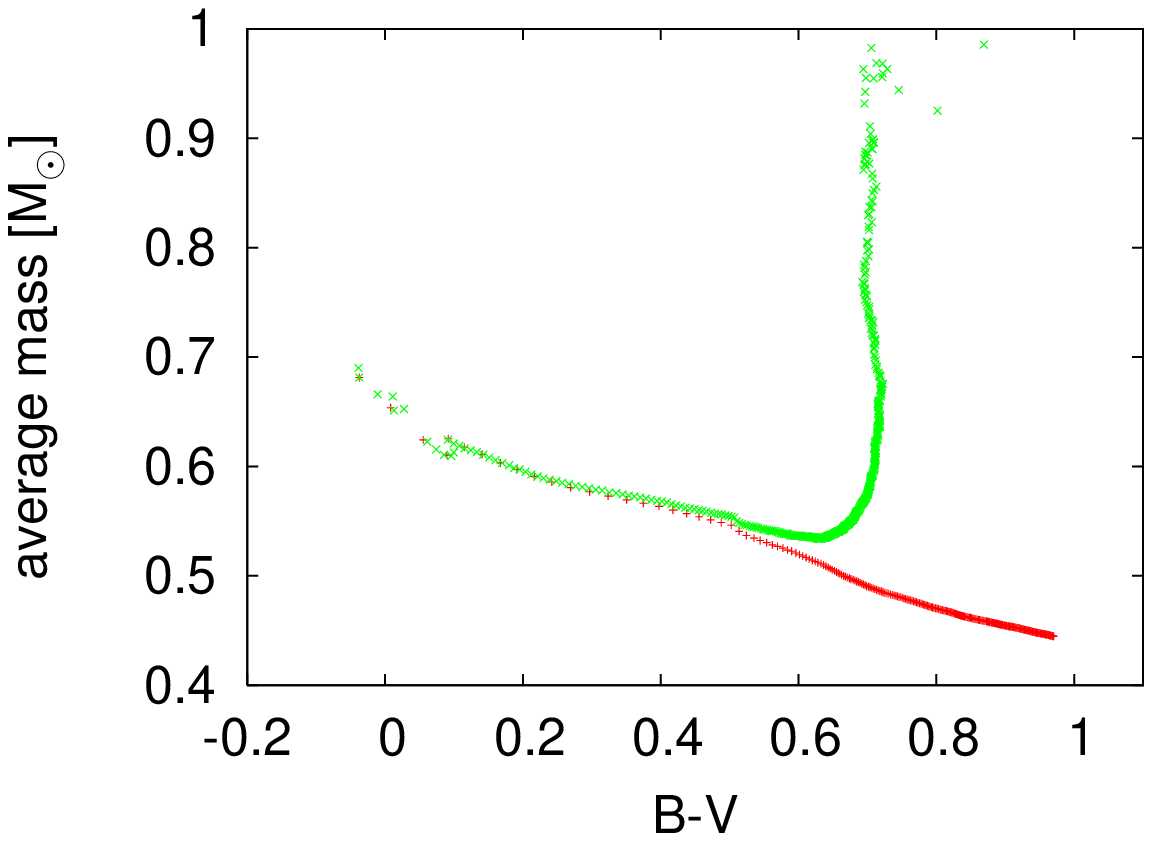}
\caption{Evolution of the average mass per star. The red line shows
the isolated cluster and the green line shows the a cluster influenced by
a tidal field with $\bar{R} = 15$. The left plot shows the time evolution, 
whereas the right plot shows the average mass evolution in terms of the B-V color.}
\end{figure*}

\section{Comparison to the observations}
The results presented in the last section can be regarded as predictions which
need to be compared with observational results. So far, we have restricted
ourselves to integrated properties such as integrated colors and integrated
M/L ratios. In order to compare these results to observations, we take the
integrated colors of galactic globular clusters from the Harris catalogue
(Harris et al. 1996). Integrated M/L ratios were taken from 
Illingworth (1976) and from Mandushev et al. (1991) by using their mass 
estimations from single-mass King models. The comparison to our predicted
color-M/L relation is shown in Figure 17 In addition, we take the M/L
ratios from Pryor \& Meylan (1993), who use a family of multi-component
King-Mitchie models to estimate their masses. The comparison of those
M/L values with our simulations is shown in Figure 18. Whereas the 
single-mass estimations of the first two authors deliver M/L ratios, which are
systematically at the lower end of our predicted M/L values, the 
Pryor \& Meylan values match well the range of M/L ratios predicted
by our simulations. In more detail, for the simulation of the isolated cluster 
a large fraction of objects still mismatches the simulated results, whereas
for the cluster in the tidal field the whole sample of objects is in a M/L
regime that matches the simulated values, when the low-M/L-tail is taken
into account. In terms of colors almost a certain fraction of objects are 
redder than the color edge predicted in the simulations.  
However, the exact value of the 
color-edge depends on the parameters of the simulation. This effect is 
demonstrated in Figure 20, which shows simulations with different 
parameter combinations. The origin for the point cloud in Figure 18 is 
therefore a mixture of different parameters for the different galactic 
globular clusters. \\

One should be aware of the fact that the mass estimations of these authors 
rely on models which
are different from the model considered here. If one were aiming at
self consistence, a better test would be to compare directly the velocity
dispersions used by these authors with those predicted by our models. 
Such direct comparison is shown in Figure 19. It shows the velocity dispersion
and the core radius of galactic globular clusters. This can be compared to
our simulated globular cluster, for which the corresponding parameters are
shown as a purple curve, which represents the time evolution of the simulated
cluster. 
From Figure 19 we conclude that the core radii of the simulated cluster are within  
the regime of those core radii occuring
for galactic globular clusters, for the entire lifetime of the simulated cluster. 
In the beginning the velocity dispersion of the simulation is in a region 
where galactic globular clusters are present,
but at the end of the simulation it becomes too low by a factor of 10.
The reason is our low particle number, which leads to a relatively low
cluster mass, in particular at higher ages, where the cluster has lost a
large fraction of its initial mass. \\

\begin{figure*}
\centering
\includegraphics[width=8cm]{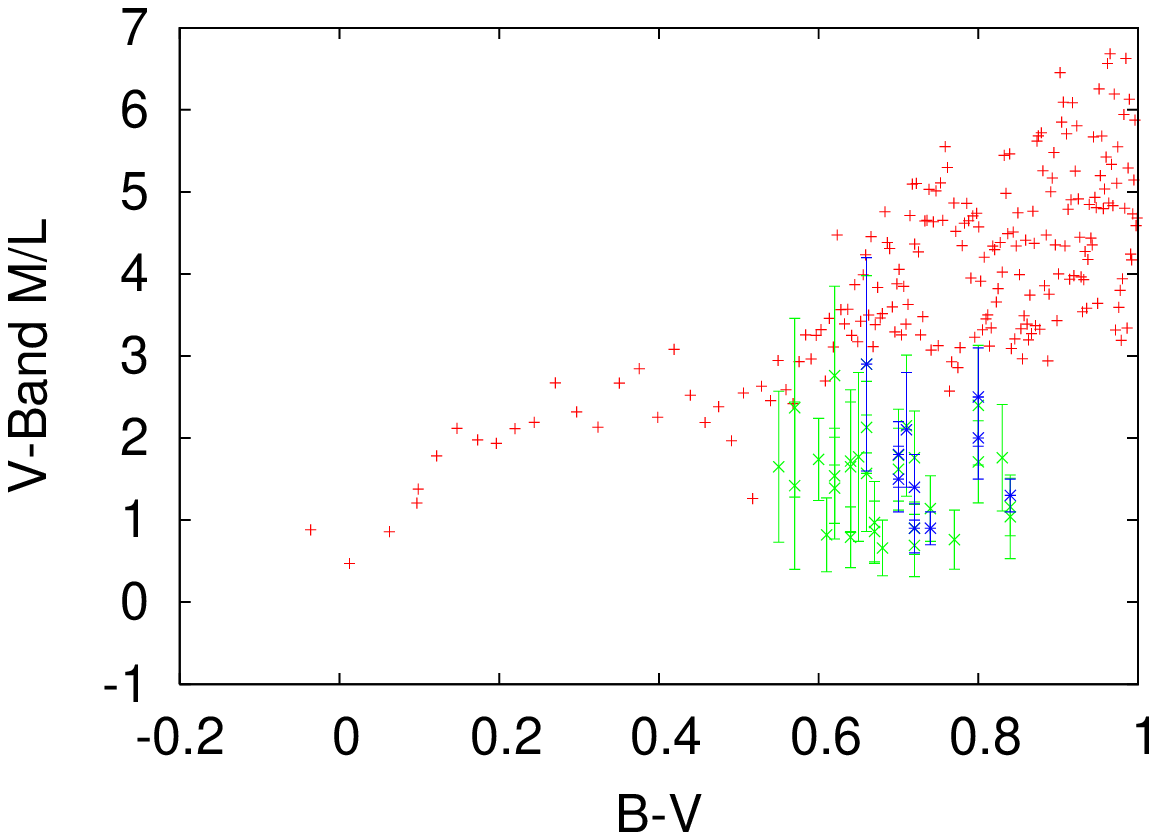}
\includegraphics[width=8cm]{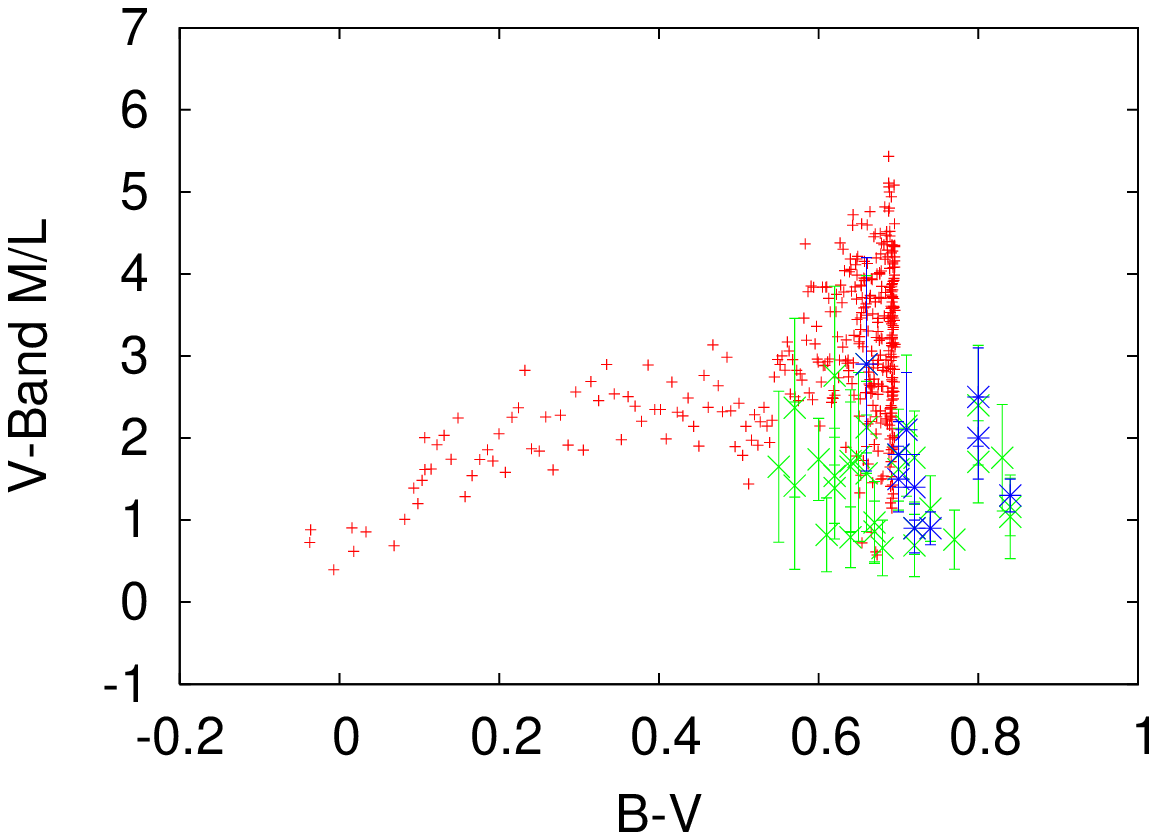}
\caption{Comparison of the measured B-V color vs. M/L ratio correlation to
our simulation results, as shown in Figure 15. The left plot shows the
simulation of a isolated cluster, whereas the right plot shows the simulation
of a cluster in a tidal field. The integrated colors of galactic globular
clusters are from the Harris catalogue (Harris 1996). The M/L ratio 
estimations are from Illingworth (1976) (blue points) and from 
Mandushev et al. (1991) (green points). From both authors we use the M/L 
values estimated by using a single mass King model.}
\end{figure*}

\begin{figure*}
\centering
\includegraphics[width=8cm]{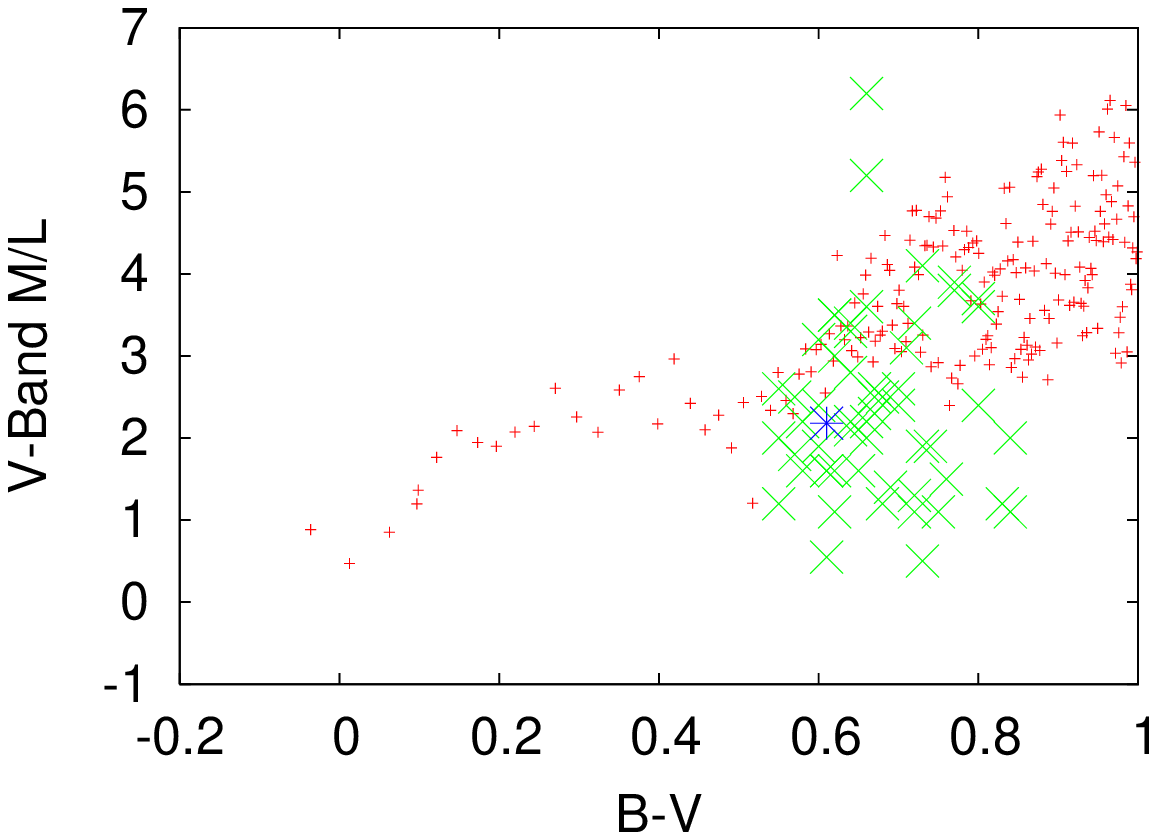}
\includegraphics[width=8cm]{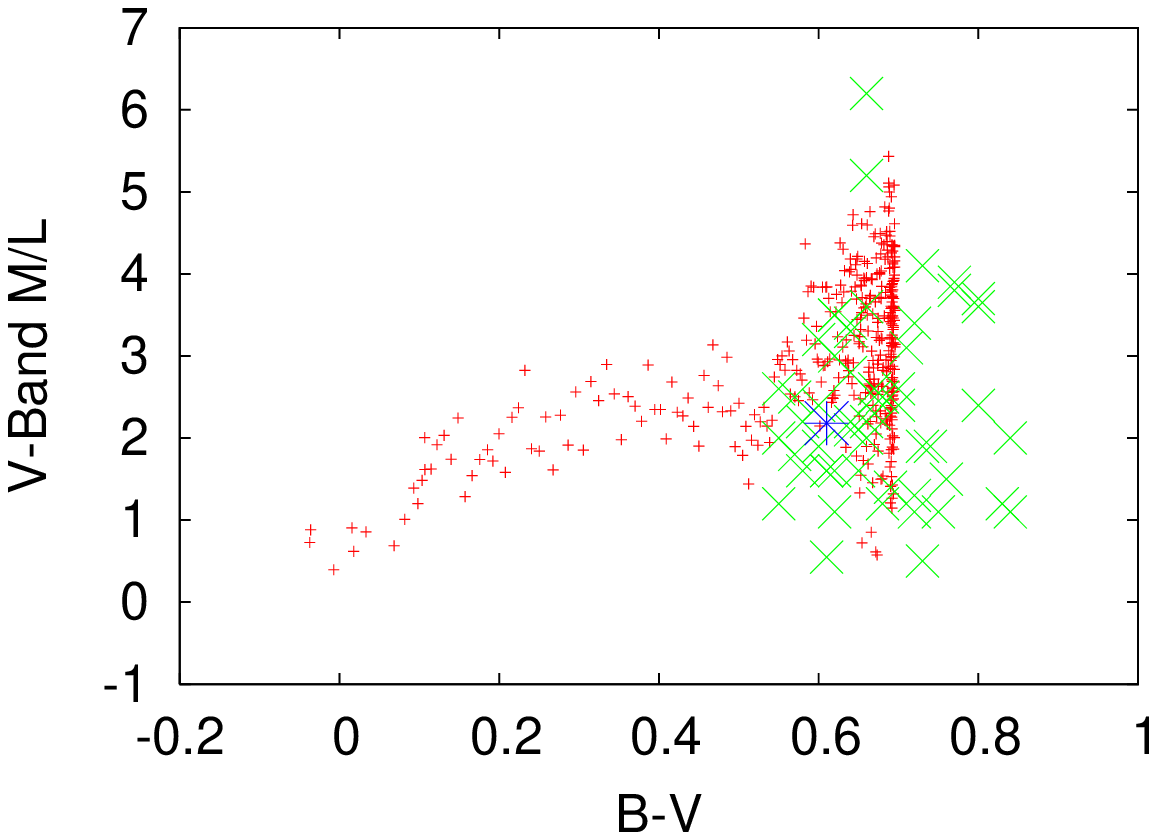}
\caption{Same as Figure 23. This time we use the M/L ratio estimations from
Pryor \& Meylan (1993), who use multi-component King-Mitchie models for their
mass estimation. The blue point is NGC 2298, for the mass we use the value
given in De Marchi \& Pulone (2006).}
\end{figure*}

\begin{figure*}
\centering
\includegraphics[width=9cm]{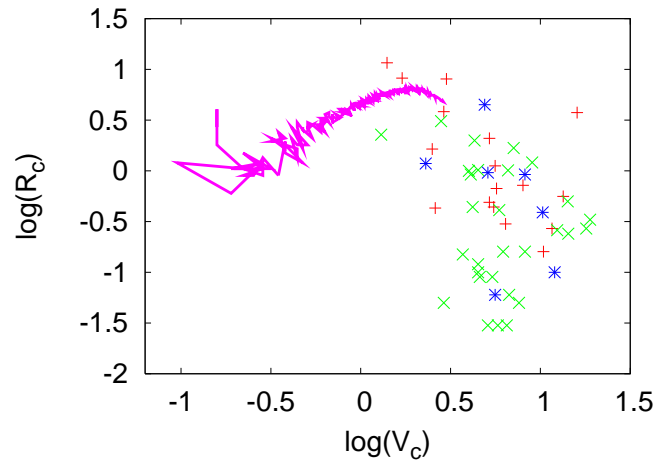}
\caption{Velocity dispersion vs. core radius for galactic globular clusters,
as listed in Pryor \& Meylan (1993), who summarized these values. The authors
who published these values are given as references in (Pryor \& Meylan 1993).
Color coded are different intervals of M/L ratios estimated by (Pryor 
\& Meylan 1993): Red: 0.5 $\le$ M/L $\le$ 1.6, green: 1.8  $\le$ M/L $\le$ 3.5,
blue: 3.6 $\le$ M/L $\le$ 6.2. As a comparison, our simulated cluster is shown
as a purple line, which represents the time evolution of these parameter
values.}
\end{figure*}

\begin{figure*}
\centering
\includegraphics[width=8cm]{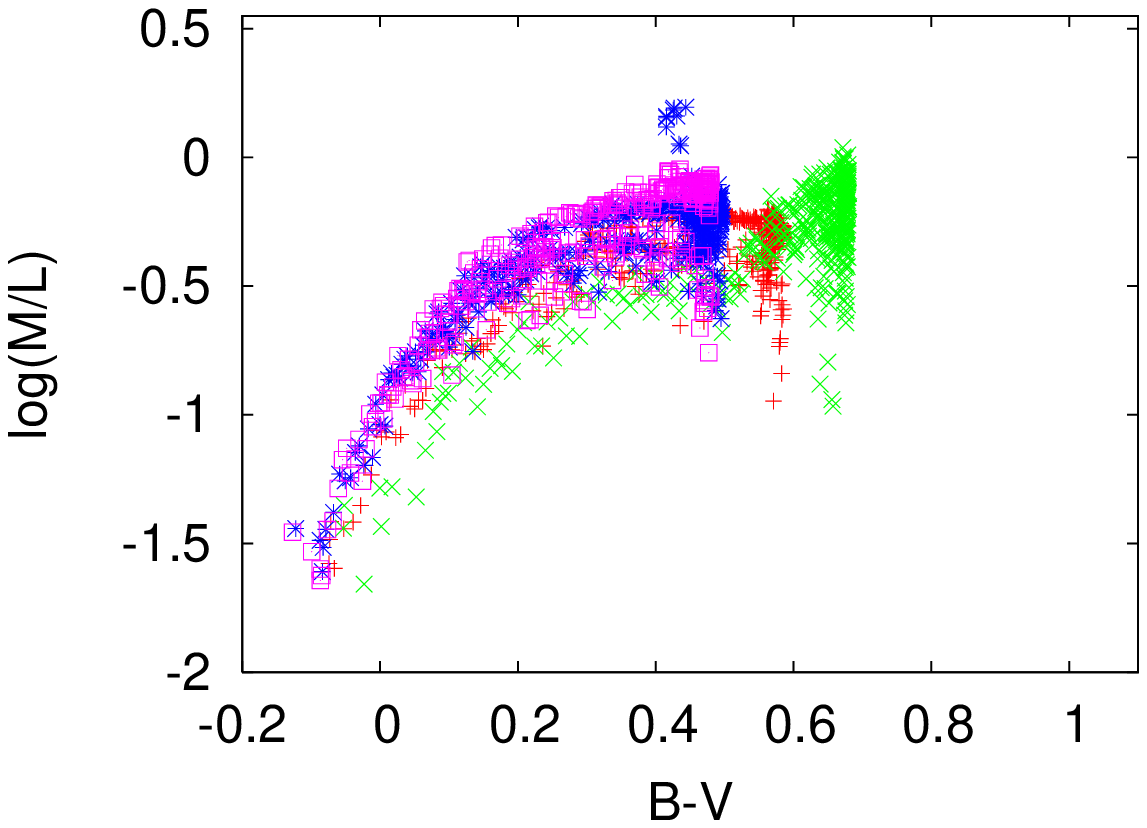}
\includegraphics[width=8cm]{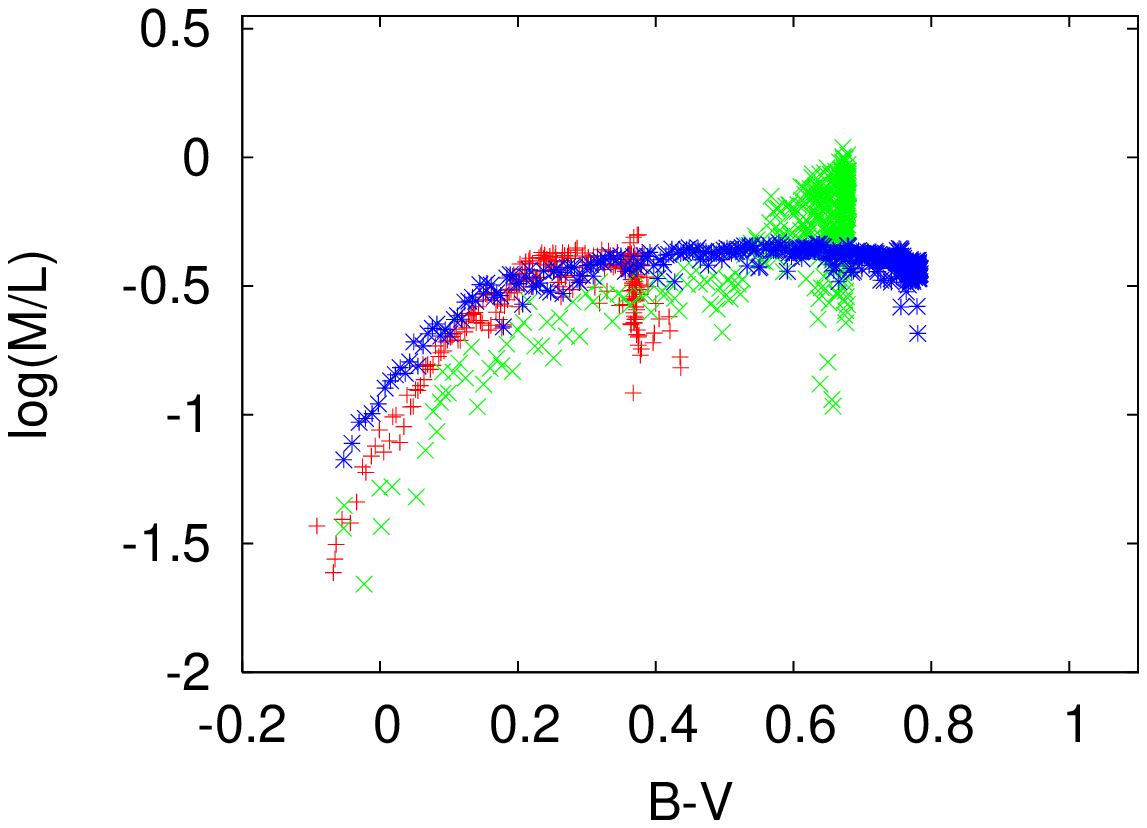}
\caption{Mass-to-light ratio vs. B-V color for clusters in a tidal field. This
 time simulations with different parameters are shown. The left figure shows
different metallicities: red: [Fe/H] = -0.7, green: [Fe/H] = -1.0, blue: [Fe/H] = -1.2, purple: [Fe/H] = -1.5. The right figure shows different galactocentric
distances: red: 4 kpc, green: 8.5 kpc, blue: 16 kpc.}
\end{figure*}

In order to compare our results in terms of spectra to the
observations, we use the library of integrated spectra of galactic globular
clusters published by Schiavon et al. (2005). This library contains 40
cluster spectra covering the range of 335 - 643 nm with 0.31 nm resolution.
The sample of globular clusters contains objects with a wide range in
parameters such as metallicity and galactocentric distance. We compared
these spectra with our simulated spectra. Since all spectra are very different,
most spectra do not match at all to our simulated one. By chance we found
one object, whose spectrum matches quite well with our simulated one.
It is NGC 1851, which has a metallicity of [Fe/H] = $-1.22$ and a 
galactocentric distance of 16.7 kpc, when looking to the Harris catalogue. 
Whereas the metallicity matches perfectly to our simulated one, the 
galactocentric distance is twice the value we simulated. 
This comparison is shown in Figure 21. \\

\begin{figure*}
\centering
\includegraphics[width=13cm]{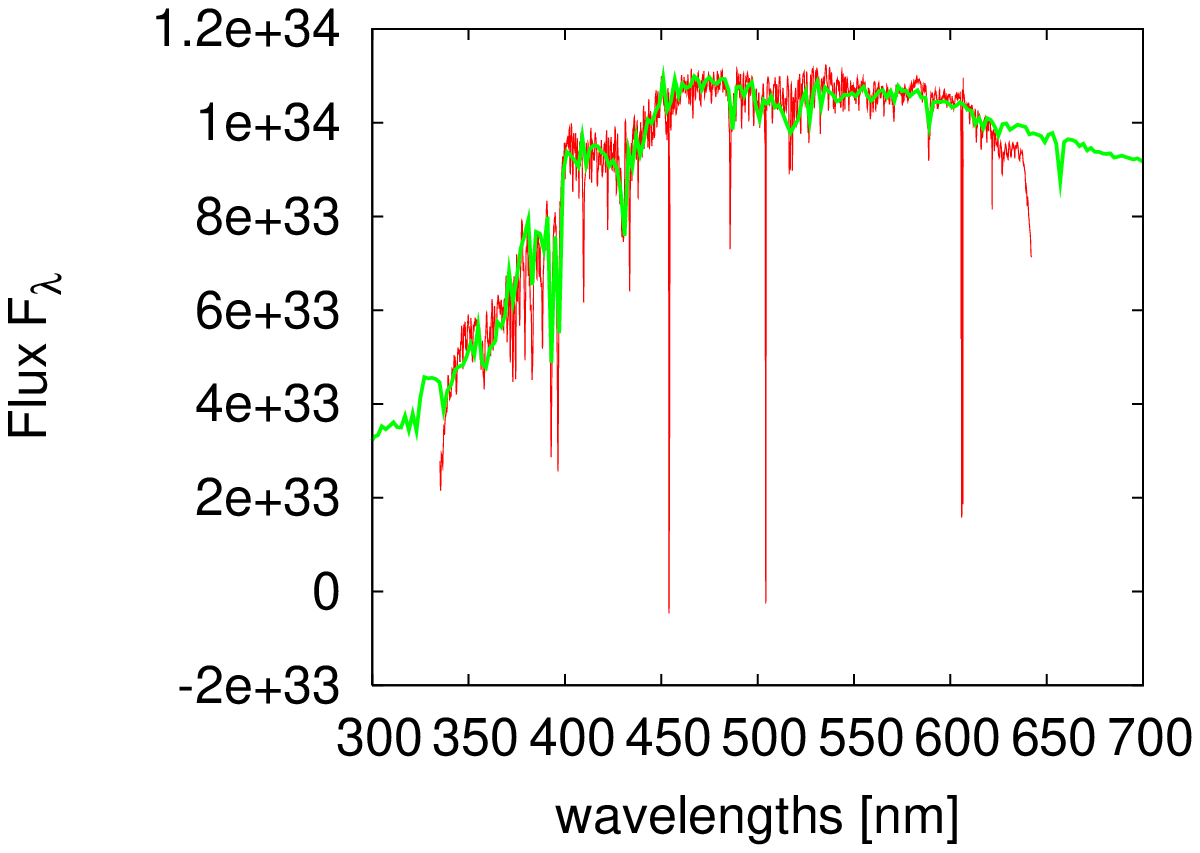}
\caption{Comparison of the integrated spectrum of NGC 1851 measured by
Schiavon et al. (2005) to our simulated spectrum of a cluster in a tidal
field.}
\end{figure*}

\begin{table}
\caption{Comparison of model parameters for M31 globular cluster SED's observed
by Ma et al. (2006 b).}
\centering
\begin{tabular}{c c c c c}
\hline
Object & \multicolumn{2}{c}{age [Gyr]} &  \multicolumn{2}{c}{metallicity [Fe/H]} \\
name & Ma et al. & this & Ma et al. & this  \\
& (2006 b) & work & (2006 b) & work \\
\hline 
M 009 & 2.92 $\pm$ 0.65 & 8 & -0.63 $\pm$ 0.06 & -1.0 \\
M 013 & 7.65 $\pm$ 1.03 & 8 & -1.10 $\pm$ 0.06 & -1.0 \\
M 024 & 9.34 $\pm$ 1.86 & 8 & -1.46 $\pm$ 0.22 & -1.0 \\
M 035 & 4.36 $\pm$ 0.44 & 8 & -1.70 $\pm$ 0.15 & -1.0 \\
\hline
\end{tabular}
\end{table}

We compare our results also to extragalactic globular clusters. Ma et al. 
published spectral energy distributions of 42 M81 globular clusters
(Ma et al 2006 a) and of 15 M31 globular clusters and 30 cluster candidates 
(Ma et al 2006 b) in 13 intermediate band filters from 400 to 1000 nm. 
We compared them to our integrated spectra by eye. 

Figure 23 shows the 4 best matching results in the M81 case. 
For these four objects the long-wavelength upturn above $\approx$
600 nm is reproduced very well. The 4 best matching results in the M31 case
are shown in Figure 24. If we neglect the first filter with a central 
wavelength of 421 nm, these four examples match the simulated spectrum
even better than in the M81 case. \\

Ma et al. (2006 b) tried to estimate ages and metallicities of the M31 
globular clusters by $\chi^2$ fitting of Bruzual-Charlot models to their
13 filter SEDs. Of course, a dircet comparison to our model does
not make sense, because our method of creating stellar populations is
entirely different from that of the Bruzual Charlot model in the case of
globular clusters in a tidal field. Another issue is the well-known
age-metallicity degeneracy. \\

However, it is interesting to see how their fit parameters relate
to our model paramters.
Whereas our simulation has in all four cases the same age and metallicity,
the Ma et al. fits span a range between 2.9 Myr and 9.3 Myr in age, and
a range between -0.6 and -1.7 in metallicity. We think that at least for the
upturn at wavelengths above 600 nm our one model fits all four spectra well. \\

\begin{figure*}
\centering
\includegraphics[width=16cm]{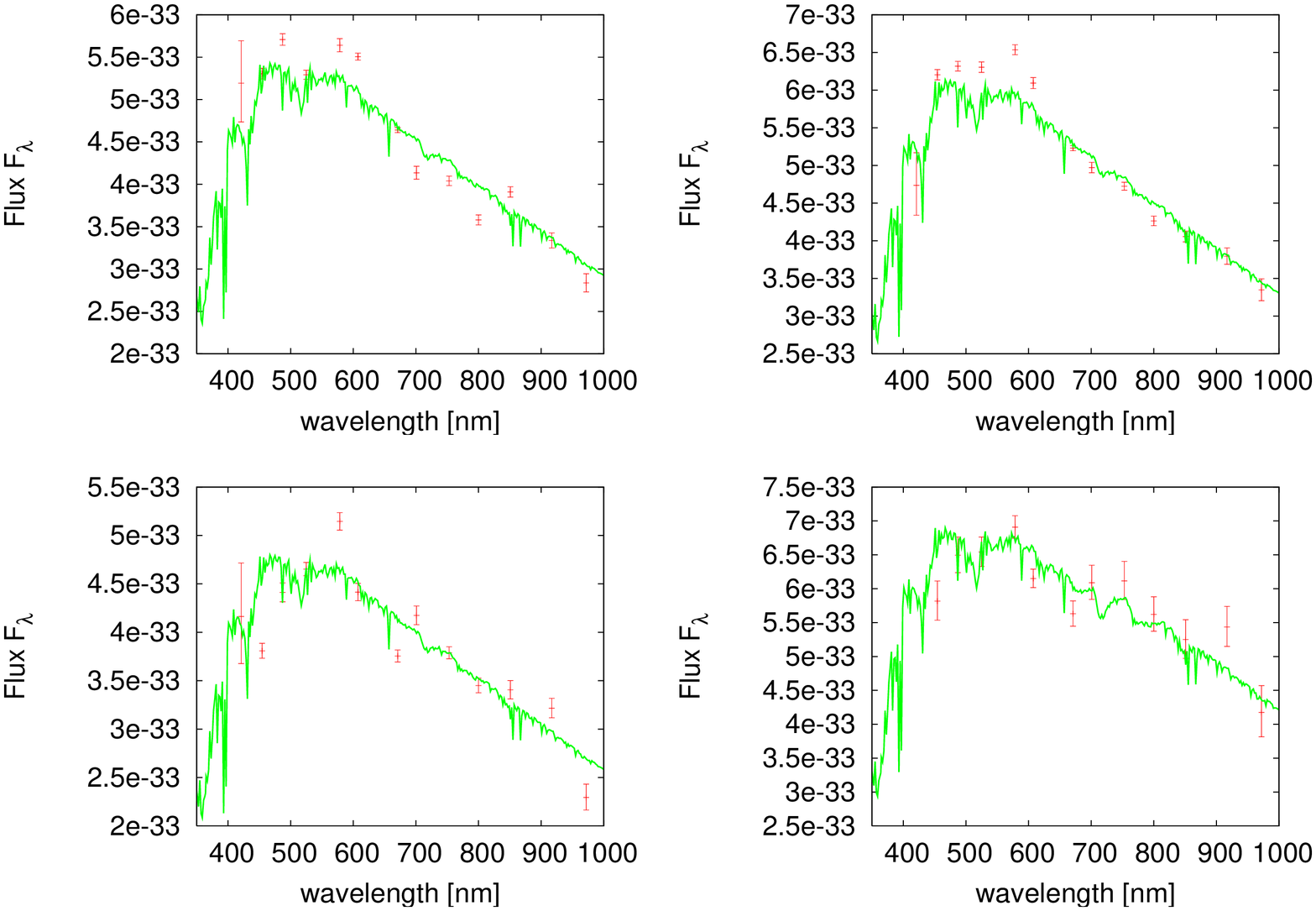}
\caption{Comparison of the SED's of M81 globular clusters measured by Ma et al.
(2006 a) to our simulated spectrum of a cluster in a tidal field. For the 
object in the left plot in the lower panel an age of 5 Gyr is suitable, for 
the other objects the spectrum with an age of 8 Gyr fits good.}
\end{figure*}

\begin{figure*}
\centering
\includegraphics[width=16cm]{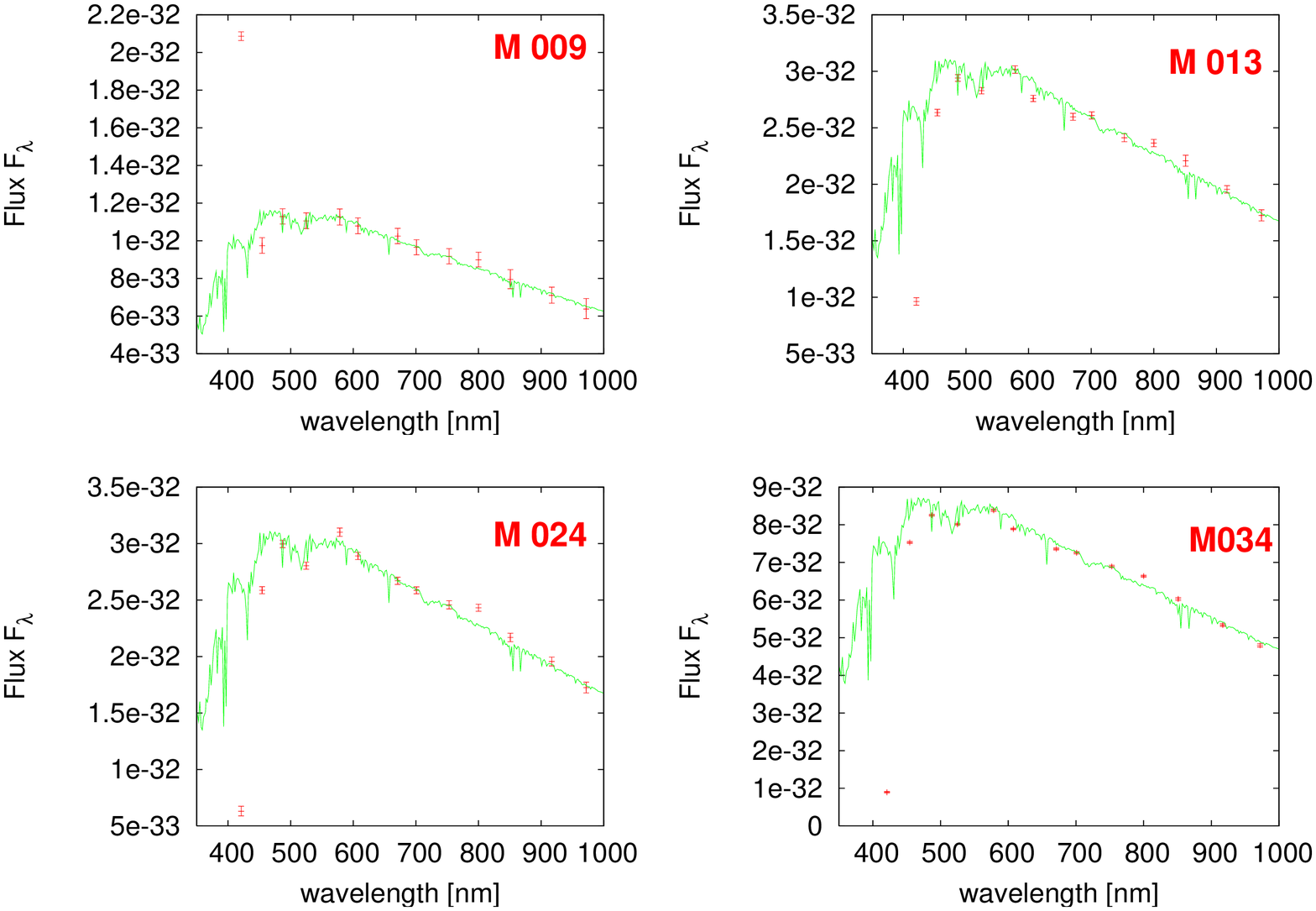}
\caption{Comparison of the SED's of M31 globular clusters measured by Ma et al.
(2006 b) to our simulated spectrum of a cluster in a tidal field. For all
four objects the spectrum at an age of 8 Gyr fits good.}
\end{figure*}

\section{Conclusion}
The comparison of the NBODY6++ results and the PEGASE results of the integrated
spectra shows an agreement which is at least promising. When comparing at 
solar metallicity, there is a good agreement at short wavelengths, reflecting
a good agreement in the U-B color at all ages. The main difference occurs
at low ages in the medium wavelength regime and at high ages in the longer
wavelength regime. When looking to the metallicity dependence the situation
becomes even more complex. \\

In both cases the BaSeL 2.0 library is used for transforming stellar parameters
into spectra and colors. The difference at wavelengths
above 700 nm, which occurs in the 15 Gyr plot, is probably due to the TP-AGB 
stars. There are no TP-AGB stars taken into account in this work. \\

For the cluster in the tidal field one would expect that the cluster becomes
bluer. Energy equipartition is expected to sort the particles by mass in the
sense that high-mass particles sink to the cluster center, whereas low-mass
particles have a higher probability to escape from the cluster after leaving
the tidal radius. Hence, one would expect that the escape of red low-mass
stars leads to a relatively bluer integrated color. Indeed, there 
exists at least a parameter combination where the cluster
becomes bluer, when the concentration is high enough and the lifetime
is long enough compared to the stellar evolution lifetime. \\

The last two results raise the question of how the tidal field
also influences the mass to light ratio. In Figure 15 the M/L ratio
is compared for an isolated cluster and for a cluster influenced by a tidal
field. The isolated cluster shows a very good correlation between color and
M/L ratio. A similar correlation was also found by Bell and de Jong in the case
of spiral galaxies (Bell \& de Jong 2001). \\

For the cluster in the tidal field we chose the $\bar{R}$=15 case (black
line in Figure 6). It shows significantly the expected bluer colors and 
lives for
several Gyr. In this case there is a significant deviation in the M/L
ratio at the red edge. Hence, the Bell \& de Jong-like correlation is no longer
true for clusters in a tidal field in the final disruption phase. Whereas 
a Bell \& de Jong-like correlation
can be used to determine the stellar mass of isolated clusters, one would
introduce considerable deviations when using it for clusters influenced by
a tidal field. \\

The comparison to the observed color-M/L ratio correlation shows a good
agreement between observations and our simulations in terms of the
V-band M/L ratios. The observed M/L ratios tend to confirm the
lower values predicted by our model of a cluster in a tidal field rather than
the model of a isolated cluster. \\

However, one should be aware that the observed M/L estimations depend also
on a certain model. In one case this is a one-component King model, in the 
other case these are  multi-component King-Mitchie models. The comparison
of these model-dependent estimations to our simulations is of course not
self-consistent. Therefore we also compared the observables themselves,
namely the central velocity dispersion and the core radius, to our
simulation, and we found values which are at least present among the
sample of galactic globular clusters. \\

In terms of colors our observed color-edge on the red side is confirmed 
by the observations for a certain fraction of objects. However, there are 
still lots of objects above our predicted color-edge, and therefore we need 
more simulations with different 
parameters. For example, a lower $\bar{R}$ value would increase the cluster
lifetime and therefore leads to a redder color edge.  \\

The overall impression of Figure 18 is that our model of the cluster in a tidal
field matches the observations of galactic globular clusters quite well.
 Since a cluster in a tidal field cannot
be modeled with ordinary stellar population synthesis models, we have here
a direct test for the validity of neglecting the effects of dynamical
evolution of globular clusters. Another interesting feature is the cluster
NGC 2298, which is the blue point in Figure 18. Recently, De Marchi \& Pulone
(2006)
published an investigation of this object and claim that this cluster is on
its way to disruption. In our comparison to the model this object lies
exactly at the predicted color edge, and the M/L ratio also fits well to
our predicted value. Therefore, this object fits nicely in our picture of a
tidally disrupted cluster. \\

Integrated colors and spectra are tailored to objects which are
too far away to be resolved into single stars. But also for galactic globular
clusters integrated colors and spectra are available, and in spite of the
existence of more detailed data we will use them in order to figure out
what we can learn by just observing integrated SEDs. \\

Therefore, we compared the integrated spectra of our simulations to observed
spectra of galactic globular clusters. In the case of the globular
cluster NGC 1851 we find a good agreement of the integrated spectrum to
our simulated spectrum. In terms of parameters, the cluster has a similar
metallicity to our simulated cluster, and the galactocentric distance is
twice the value in our simulation. \\

After demonstrating the power of integrated spectra in the case of
galactic globular clusters we applied it to extragalactic objects, namely
the globular cluster systems of M81 and M31, for which observed medium
band SEDs are available in the literature. \\

For these SEDs we find a good agreement between our 
simulation and several M81 globular cluster SEDs. In the case of the
M31 globular cluster SEDs the agreement is even better. We conclude from
this, that our simulations lead to results, with are at least present among
the globular cluster systems of these two galaxies. \\

\section{Outlook} 
\subsection{Detailed parameter study}
So far, we have demonstated, that for simulations with certain parameters
the results are able to to explain observations of some globular clusters.
In future, more simulations will allow us to cover broader regions of the
parameter space. $\chi^2$ fitting techniques will show in more
detail possible parameter degeneracies and therefore the explanatory power
of integrated spectra of distant globular clusters. 

\subsection{Mixed stellar populations}
All simulations presented so far in this work are done with single stellar 
populations in the sense that all stars have the same age and the same 
metallicity. However, in future we are planning to deal also with mixed
stellar populations. This means we want to allow for different stellar
populations with different ages and different metallicities in the same 
simulation. As a first step towards a mixed stellar population we start
with two different populations with different metallicities. \\

Our approach is realized in the code in such a way that every second star
in a simulation has one metallicity $Z_1$ and the other stars have another
metallicity $Z_2$. All these stars are distributed over a Plummer Sphere
and are therefore ideally mixed. In future one may also think about a initial
spatial distribution of different metallicities, for example a merger of
two globular clusters with different metallicities. \\

Figure 26 shows our result of a simulation of a isolated cluster with two
different metallicities. For large ages the B-V colors are exactly in between
the colors of a single stellar population of one of these two metallicities.
We conclude that these result is very promising and we will further go into
the direction of mixed stellar polulations. \\

\begin{figure*}
\centering
\includegraphics[width=16cm]{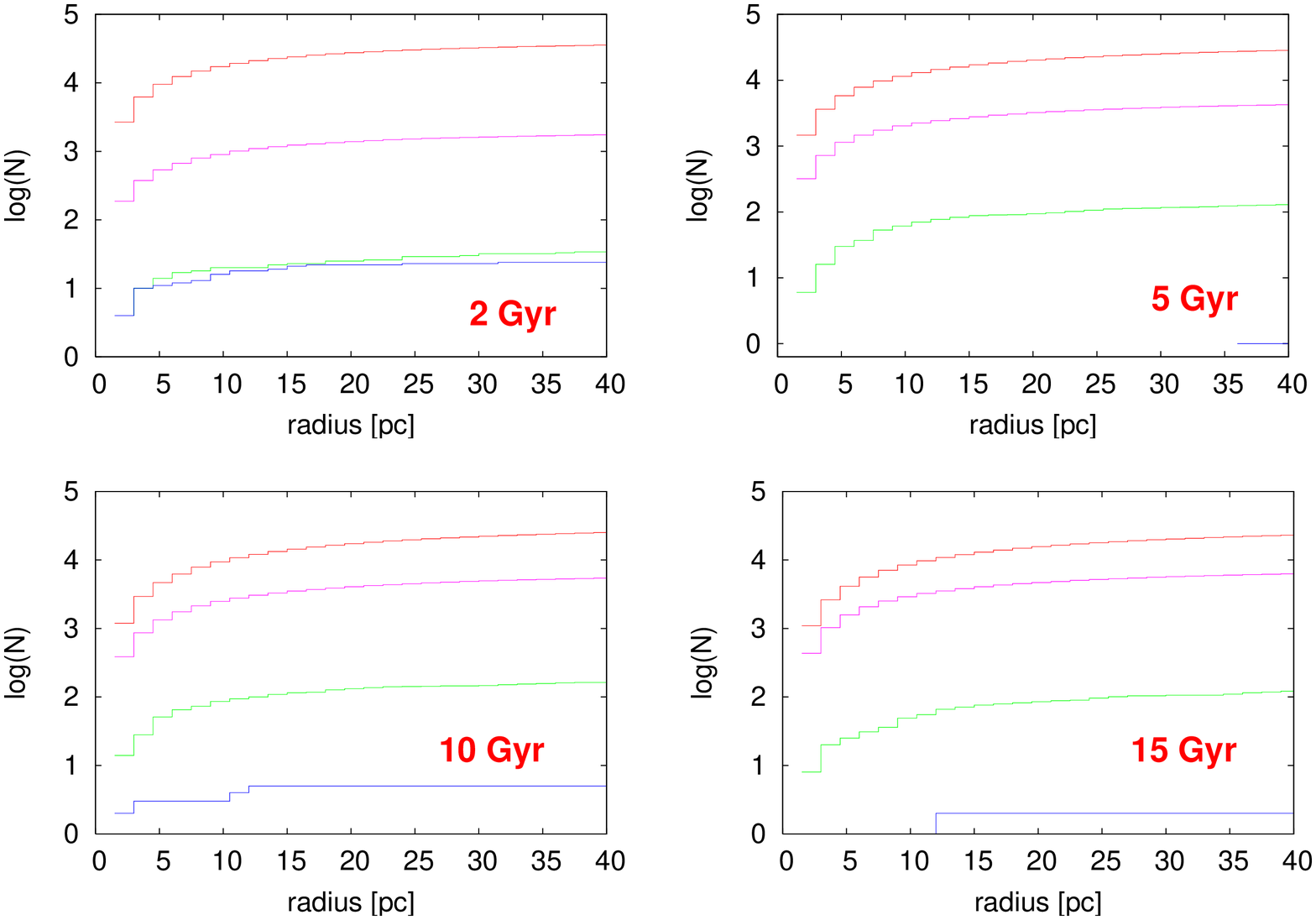}
\caption{Radial distribution of stars. The different colors are different 
stellar types. Red: main sequence stars, green: red giants, blue: AGB stars,
purple: white dwarfs.}
\end{figure*}

\begin{figure*}
\centering
\includegraphics[width=16cm]{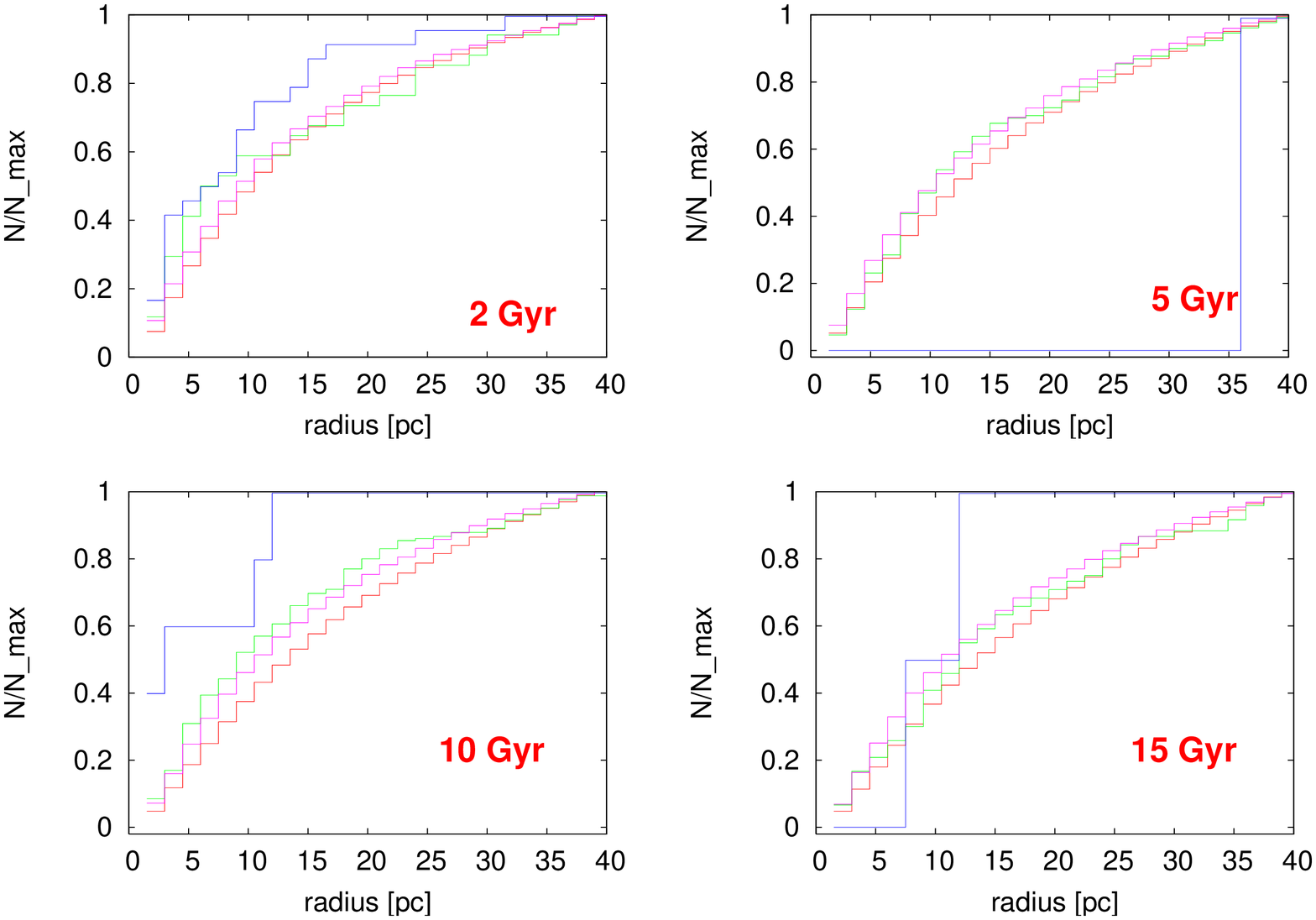}
\caption{Radial distribution of stars, normalized to 1. The different colors 
are different stellar types. Red: main sequence stars, green: red giants, 
blue: AGB stars, purple: white dwarfs.}
\end{figure*}

\begin{figure*}
\centering
\includegraphics[width=8cm]{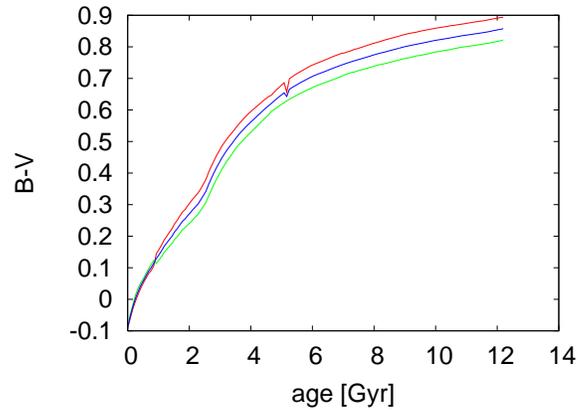}
\caption{Mixed stellar populations with two different metallicities.
The blue curve shows a simulation of a globular cluster 
containing two stellar populations with different metallicities:
$Z_1 = Z_{\odot}$ and $Z_2 = \frac{1}{10} Z_{\odot}$.  As a 
comparison, in the red and green curves only particles of the same metallicity
are plotted. 
populations with the red curve being the upper metalllicity $Z_1$ and the 
green curve being the lower metallicity $Z_2$.}
\end{figure*}

\end{document}